%% file: judging-judges-gymnastics.tex
\documentclass[twocolumn,conference]{IEEEtran}

\usepackage[style=ieee,backend=bibtex,sortcites=true,sorting=nyt,hyperref=false,backref=false]{biblatex}
\usepackage{microtype}

\usepackage{amssymb}
\usepackage{amsmath}
\usepackage{amsfonts}
\usepackage{bbm}
\usepackage{graphicx}

\usepackage{hyperref}
\usepackage{float}
\usepackage{placeins}

\usepackage{subcaption}
\usepackage{url}
\usepackage{flushend}

\usepackage{pgfplotstable}
\usepackage{booktabs}

\newtheorem{theorem}{Theorem}[section]

\newtheorem{defn}[theorem]{Definition}

\usepackage{xspace}
\newcommand{\jep}{\text{JEP}\xspace}

\usepackage[T1]{fontenc}
\usepackage[utf8]{inputenc}
\usepackage[english]{babel}

\usepackage{float}

\usepackage{array}
\usepackage[export]{adjustbox}
\usepackage{fancyhdr}

\usepackage{ifthen}
\usepackage{xcolor}

\newboolean{showcomments}
\setboolean{showcomments}{true}
\ifthenelse{\boolean{showcomments}}
{ \newcommand{\mynote}[3]{
    \fbox{\bfseries\sffamily\scriptsize#1}
    {\small$\blacktriangleright$\textsf{\emph{\color{#3}{#2}}}$\blacktriangleleft$}}}
{ \newcommand{\mynote}[3]{}}

\graphicspath{{figures/}}
\bibliography{bib-sports}

\begin{document}

\newpage

\title{Judging the Judges: Evaluating the Performance \\ of International Gymnastics Judges}

\author{
\IEEEauthorblockN{Hugues Mercier}
\IEEEauthorblockA{Universit\'{e} de Neuch\^{a}tel \\ Switzerland \\ hugues.mercier@unine.ch}
\and
\IEEEauthorblockN{Sandro Heiniger}
\IEEEauthorblockA{Universität St.Gallen \\ Switzerland \\ sandro.heiniger@unisg.ch}
}

\maketitle

\input{abstract}

\input{introduction}

\input{data}

\input{controlscore}

\input{marking}

\input{outlier}

\input{ranking}
\input{observations}

\input{conclusion}

\input{acknowledgements}

\printbibliography

\end{document}

%% file: abstract.tex
\begin{abstract}

Judging a gymnastics routine is a noisy process, and the performance of judges varies widely. In this work, we design, describe and implement a statistical engine to analyze the performance of gymnastics judges during and after major competitions like the Olympic Games and the World Championships. The engine, called the Judge Evaluation Program (JEP), has three objectives: (1) provide constructive feedback to judges, executive committees and national federations; (2) assign the best judges to the most important competitions; (3) detect bias and outright cheating. 
	
Using data from international gymnastics competitions held during the 2013--2016 Olympic cycle, we first develop a \emph{marking score} evaluating the accuracy of the marks given by gymnastics judges. Judging a gymnastics routine is a random process, and we can model this process very accurately using heteroscedastic random variables. The marking score scales the difference between the mark of a judge and the theoretical performance of a gymnast as a function of the intrinsic judging error variability estimated from data for each apparatus. This dependence between judging variability and performance quality has never been properly studied. We then study \emph{ranking scores} assessing to what extent judges rate gymnasts in the correct order, and explain why we ultimately chose not to implement them. We also study outlier detection to pinpoint gymnasts who were poorly evaluated by judges. Finally, we discuss interesting observations and discoveries that led to recommendations and rule changes at the Fédération Internationale de Gymnastique (FIG).
	
\end{abstract}


%% file: introduction.tex
\section{Introduction}

Gymnastic judges and judges from similar sports are susceptible to well-studied biases\footnote{Consult \textcite{Lan1970} for an initial comprehensive survey until 1970, and \textcite{Bar-Eli:2011} for a recent survey.}. \textcite{Ansorge:1988} detected a \emph{national bias} of artistic gymnastics judges at the 1984 Olympic Games: judges tend to give  better marks to athletes from their home country while penalizing close competitors from other countries. National bias was subsequently detected in rhythmic gymnastics at the 2000 Olympic Games \cite{Popovic:2000}, 
and in numerous other sports such as figure skating \cite{Campbell:1996, Zitzewitz:2006}, Muay Thai boxing \cite{Myers:2006}, ski jumping \cite{Zitzewitz:2006}, diving \cite{Emerson:2009} and dressage \cite{Sandberg:2018}.

\textcite{Plessner:1999} observed a \emph{serial position bias} in gymnastics experiments: a competitor performing and evaluated last gets better marks than when performing first. \textcite{Boenetal:2008} found a \emph{conformity bias} in gymnastics: open feedback causes judges to adapt their marks to those of the other judges of the panel. \textcite{Damisch:2006} found a \emph{sequential bias} in artistic gymnastics at the 2004 Olympic Games: the evaluation of a gymnast is likely more generous than expected if the preceding gymnast performed well. \textcite{Plessner:2005} showed in an experiment that still rings judges can make systematic errors based on their viewpoint. Biases observed in other sports might also occur in gymnastics as well. \textcite{Findlay:2004} found a \emph{reputation bias} in figure skating: judges overestimate the performance of athletes with a good reputation. \textcite{Price:2010} quantified the \emph{racial bias} of NBA officials against players of the opposite race, which was large enough to affect the outcome of basketball games. Interestingly, the racial bias of NBA officials subsequently disappeared, most probably due to the public awareness of the bias from the first study \cite{PPW2013}.

The aforementioned biases are often unconscious and cannot always be entirely eliminated in practice. However, rule changes and monitoring from the Fédération Internationale de Gymnastique (FIG) as well as increased scrutiny induced by the media exposure of major gymnastics competitions make these biases reasonably small and tempered by mark aggregation. In fact, judging is much more about skill and training than bias: it is difficult to evaluate every single aspect of the complex movements that are part of a gymnastics routine, and unsurprisingly nearly all international judges are former gymnasts. This challenge has been known since at least the 1930s \cite{Zwarg1935}, and there is a large number of studies on the ability of judges to detect execution mistakes in gymnastic routines \cite{Ste-Marie:1999,Ste-Marie:2000,Pizzera:2012,Flessas:2015,Pizzera:2018}\footnote{Consult \textcite{Lan1970} for an initial comprehensive survey until 1970, and \textcite{Bar-Eli:2011} for a recent survey.}. In a nutshell, novice judges consult their scoring sheet much more often than experienced international judges, thus missing execution errors. Furthermore, international judges have superior perceptual anticipation, are better to detect errors in their peripheral vision and, when they are former gymnasts, leverage their own sensorimotor experiences.

Even among well-trained judges at the international level, there are significant differences: some judges are simply better than others. For this reason, the FIG has developed and used the Judge Evaluation Program (JEP) to assess the performance of judges during and after international competitions. The work on JEP was started in 2006 and the tool has grown iteratively since then. Despite its usefulness, JEP was partly designed with unsound and inaccurate mathematical tools, and was not always evaluating what it ought to evaluate.

\subsection{Our contributions}

In this article, we design and describe a toolbox to assess, as objectively as possible, the accuracy of international gymnastics judges using simple yet rigorous tools. 
This toolbox is now the core statistical engine of the new iteration of JEP\footnote{The new iteration of JEP was developed in collaboration with the FIG and the Longines watchmaker. It is a full software stack that handles all the interactions between the databases, our statistical engine, and a user-friendly front-end to generate statistics, recommendations and judging reports. 
} providing feedback to judges, executive committees and national federations.
It is used to reward the best judges by selecting them to the most important competitions such as the Olympic Games. It finds judges performing below expectations so that corrective measures can be undertaken.   
It provides hints about inconsistencies and confusing items in the Codes of Points detailing how to evaluate each apparatus, as well as weaknesses in training and accreditation processes.
In uncommon but important circumstances, it can uncover biased and cheating judges. 

The main tool we develop is a \emph{marking score} evaluating the accuracy of the marks given by a judge. We design the marking score such that it is unbiased with the apparatus/discipline under evaluation, and unbiased with respect to the skill level of the gymnasts. In other words, the main difficulty we overcome is as follows: a parallel bars judge giving 5.3 to a gymnast deserving 5.0 must be evaluated more generously than a vault judge giving 9.9 to a gymnast deserving 9.6, but how much more? To quantify this, we model the behavior of judges as heteroscedastic random variables using data from international and continental gymnastics competitions held during the 2013--2016 Olympic cycle. The standard deviation of these random variables, describing the \emph{intrinsic judging error variability} of each discipline, decreases as the performance of the gymnasts improves, which allows us to quantify precisely how judges compare to their peers.
To the best of our knowledge, this dependence between judging variability and performance quality has never been properly studied in any setting (sport or other). 

Besides allowing us to distinguish between accurate and erratic judges, we also use the marking score as the basic tool to detect outlier evaluations. The more accurate is a judge, the lower is his/her outlier detection threshold. 

We then study \emph{ranking scores} quantifying to what extent judges rank gymnasts in the correct order. We analyzed different metrics to compare distances between rankings such as the generalized version of Kendall’s $\tau$ distance \cite{Kumar:2010}. Depending on how these rankings scores are parametrized, they are either unfair by penalizing unlucky judges who blink at the wrong time, or correlated with our marking score and thus unnecessary. Since no approach was satisfactory, the FIG no longer uses ranks to monitor its judges\footnote{The previous iteration of \jep used a rudimentary ranking score.}. 

We made other interesting observations that led to recommendations and changes at the FIG during the course of this work. We show that so called reference judges, hand-picked by the FIG and imparted with more power than regular panel judges, are not better than these regular panel judges in the aggregate. We thus recommended that the FIG stops granting more power to reference judges. We also show that women judges are significantly more accurate than men judges in artistic gymnastics and in trampoline, which has training and evaluation implications. 

This is the first of a series of three articles on sports judging. In the second article~\cite{HM2018:nationalbias}, we refine national bias studies in gymnastics using the heteroscedastic behavior of the judging error of gymnastics judges. In the third article~\cite{HM2018:heteroscedasticity}, we show that this heteroscedastic judging error appears with a similar shape in other sports where panels of judges evaluate athletes objectively within a finite marking range.

The remainder of this article is organized as follows. We present our dataset and describe the gymnastic judging system in Section~\ref{sec:data}. We the discuss true performance quality and control scores in gymnastics in Section~\ref{sec:controlscore}. 
We derive the marking score in Section~\ref{sec:mark}. In Section~\ref{sec:outlier}, we use the marking score to detect outliers. Section~\ref{sec:rank} discusses ranking scores and why we ultimately left them aside. We present interesting observations and discoveries in Section~\ref{sec:obs} and conclude in Section~\ref{sec:conclusion} by discussing the strengths and limitations of our approach.

%% file: data.tex
\section{Data and judging in gymnastics}
\label{sec:data}

\begin{table*}
	\centering
	\begin{tabular}{lcccc}
		\toprule		
		&Typical panel  & Number of & Number \\
		Discipline & composition  & performances & of marks \\ 
		\midrule
		Acrobatic gymnastics & 4 E + 2 R & 756 & 4'870\\
		Aerobic gymnastics & 4 E + 2 R & 938 & 6'072 \\
		Artistic gymnastics  & 5 E + 2 R & 11'940 & 78'696  \\
		Rhythmic gymnastics & 5 E + 2 R & 2'841 & 19'052\\
		Trampoline 	& 5 E & 1'986 & 9'654 \\
		\bottomrule
	\end{tabular}
	\caption{Standard composition of the execution panel, number of performances and number of marks per discipline. \mbox{E = Execution judges;} R = Reference judges.}
	\label{tab:noJ}
\end{table*}

Gymnasts at the international level are evaluated by panels of judges for the difficulty, execution, and artistry components of their performances. The marks given by the judges are aggregated to generate the final scores and rankings of the gymnasts. The number of judges for each component and the aggregation method are specific to each discipline. In this article, we analyze the execution component of all the gymnastics disciplines: artistic gymnastics, acrobatic gymnastics, aerobic gymnastics, rhythmic gymnastics, and trampoline. We also evaluate artistry judges in acrobatic and aerobic gymnastics, but exclude difficulty judges from our analysis. Our dataset encompasses 21 international and continental competitions held during the 2013--2016 Olympic cycle culminating with the 2016 Rio Olympic Games. 

The execution of a gymnastics routine is evaluated by a panel of judges. Table~\ref{tab:noJ} summarizes the composition of the typical execution panel for each discipline\footnote{The execution panels do not always follow this typical composition: the qualifying phases in artistic and rhythmic gymnastics may include four execution judges instead of five, World Cup events and continental championships do not always feature reference judges, and aerobic and acrobatic gymnastics competitions can have larger execution panels.}. With the exception of trampoline, these panels include execution and reference judges. Execution and reference judges have different power and are selected differently, but they all judge the execution of the routines under the same conditions and using the same criteria. 

After the completion of a routine, each execution panel judge evaluates the performance by giving it a score between 0 and 10. Table~\ref{tab:noJ} includes the number of performances and judging marks per discipline in our dataset. The number of performances in an event is not always equal to the number of gymnasts. For instance, gymnasts who wish to qualify for the vault apparatus finals jump twice, each jump counting as a distinct performance in our analysis. The number of judging marks depends on the number of performances and the size of the judging panels.

%% file: controlscore.tex
\section{True performance quality and control scores in gymnastics}
\label{sec:controlscore}

The execution evaluation of a gymnastics routine is based on deductions precisely defined in the Code of Points of each apparatus\footnote{The 2017--2020 Codes of Points, their appendices and other documents related to rules for all the gymnastics disciplines are publicly available at https://www.gymnastics.sport/site/rules/rules.php. Competitions in our dataset were ruled by the 2013--2016 Codes of Points.}. The score of each judge can thus be compared to the theoretical \emph{true} performance of the gymnast.

In practice the true performance level is unknown, and the FIG typically derives \emph{control scores} with outside judging panels and video reviews post-competition. Unfortunately, the FIG does not provide accurate control scores for every performance: the number of control scores and how they are obtained depends on the discipline and competition. Besides, even when a control score is available, the codes of points might be ambiguous or the quality of a performance element may land between two discrete values. This still results in an approximation of the true performance, albeit a very good one. Control scores derived post-competition can also be biased, for instance if people deriving them know who the panel judges are, and what marks they initially gave.
For all these reasons, in our analysis, we train our model using the median judging mark of each performance as the control score. Whenever marks by reference judges, superior juries and post-competition reviews are available, we include them with execution panel judges and take the median mark over this enlarged panel, henceforth increasing the accuracy of our proxy of the true performance quality. We discuss the implications of training our data with the median, and control scores in general, in Section~\ref{sec:conclusion}.

%% file: marking.tex
\section{Marking score} \label{sec:mark}

We now derive a \emph{marking score} to evaluate the performance of gymnastics judges. We first describe our general approach using artistic gymnastics data in Section~\ref{subsection:artistic} and present results for the other gymnastics disciplines in Section~\ref{subsection:other}. Table~\ref{tab:notation} summarizes the notation we use in this section.

\subsection{General approach applied to artistic gymnastics}
\label{subsection:artistic}

The marking score must have the following properties. First, it must not depend on the skill level of the gymnasts evaluated: a judge should not be penalized nor advantaged if he judges an Olympic final with the world's best 8 gymnasts as opposed to a preliminary round with 200 gymnasts. Second, it must allow judges comparisons across apparatus, disciplines, and competitions. The marking score of a judge is thus based on three parameters:
\begin{enumerate}
	\item{The control scores of the performances}
	\item{The marks given by the judge}
	\item{The apparatus / discipline}
\end{enumerate}

\begin{table}
	\centering
	\begin{tabular}{ll}
		\toprule
		$p$ & Performance $p$  \\
		$\lambda_p$ & True quality level of Performance $p$ \\
		$c_p$ & Control score of Performance $p$ \\ 
		$j$ & Judge $j$ \\
		$s_{p,j}$ & Mark of Judge $j$ for Performance $p$ \\
		$\hat{e}_{p,j}$ & Judging discrepancy $s_{p,j} - c_p$ (approximates the judging error) \\ 
		& \qquad of Judge $j$ for Performance $p$ \\ 
		$m_{p,j}$ & Marking score for Performance $p$ by Judge $j$ \\
		$M_{j}$ & Marking score of Judge $j$ \\
		$d$ & Apparatus / Discipline $d$ \\
		$\hat{\sigma}_d(c_p)$ & Intrinsic judging error variability of Discipline $d$ \\
		$\alpha_d, \beta_d, \gamma_d$ & Parameters of Discipline $d$ \\
		$n$ & Number of performances in an event\\
		\bottomrule
	\end{tabular}
	\caption{Notation.}
	\label{tab:notation}
\end{table}

Let $s_{p, j}$ be the mark of Judge~$j$ for Performance $p$, and let $\hat{e}_{p,j}~\triangleq~s_{p, j}~-~c_p$ be the \emph{judging discrepancy} of Judge $j$ for Performance $p$. Since we use the median of the enlarged judging panel as the control score ($c_p \triangleq \underset{j}{\text{med}}(s_{p,j})$), thus as a proxy of the true performance level $\lambda_p$, it follows that $\hat{e}_{p,j}$ is a proxy of the \emph{judging error} of Judge $j$ for Performance~$p$. We emphasize once more that we discuss the advantages and drawbacks of using the median as control score in Section~\ref{sec:conclusion}.

\begin{figure}[h!]
	\centering
	\includegraphics[width=\columnwidth]{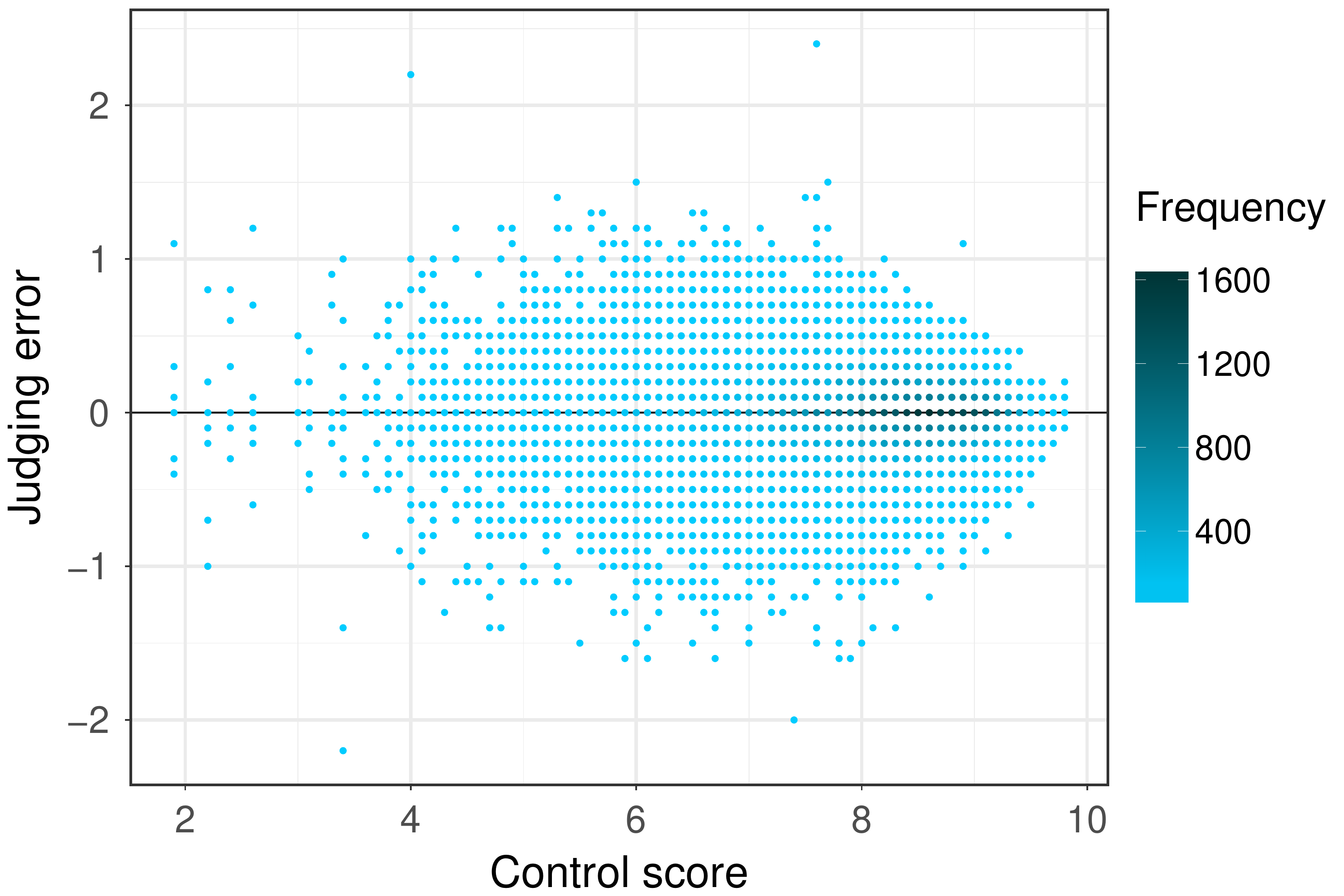}
	\caption{Distribution of the judging errors in artistic gymnastics. To improve the visibility, we aggregate the points on a $0.1 \times 0.1$ grid.}
	\label{fig:D:ag}
\end{figure}

Figure~\ref{fig:D:ag} shows the distribution of $\hat{e}_{p, j}$ for artistic gymnastics.
Our first observation is that judges are too severe as often as they are too generous, which is trivially true because we use the median as control score. The second observation is that the judging error is highly heteroscedastic. Judges are much more accurate for the best performances, and simply using $\hat{e}_{p,j}$ underweights errors made for the best gymnasts. 

\begin{figure}[h!]
	\centering
	\includegraphics[width=\columnwidth]{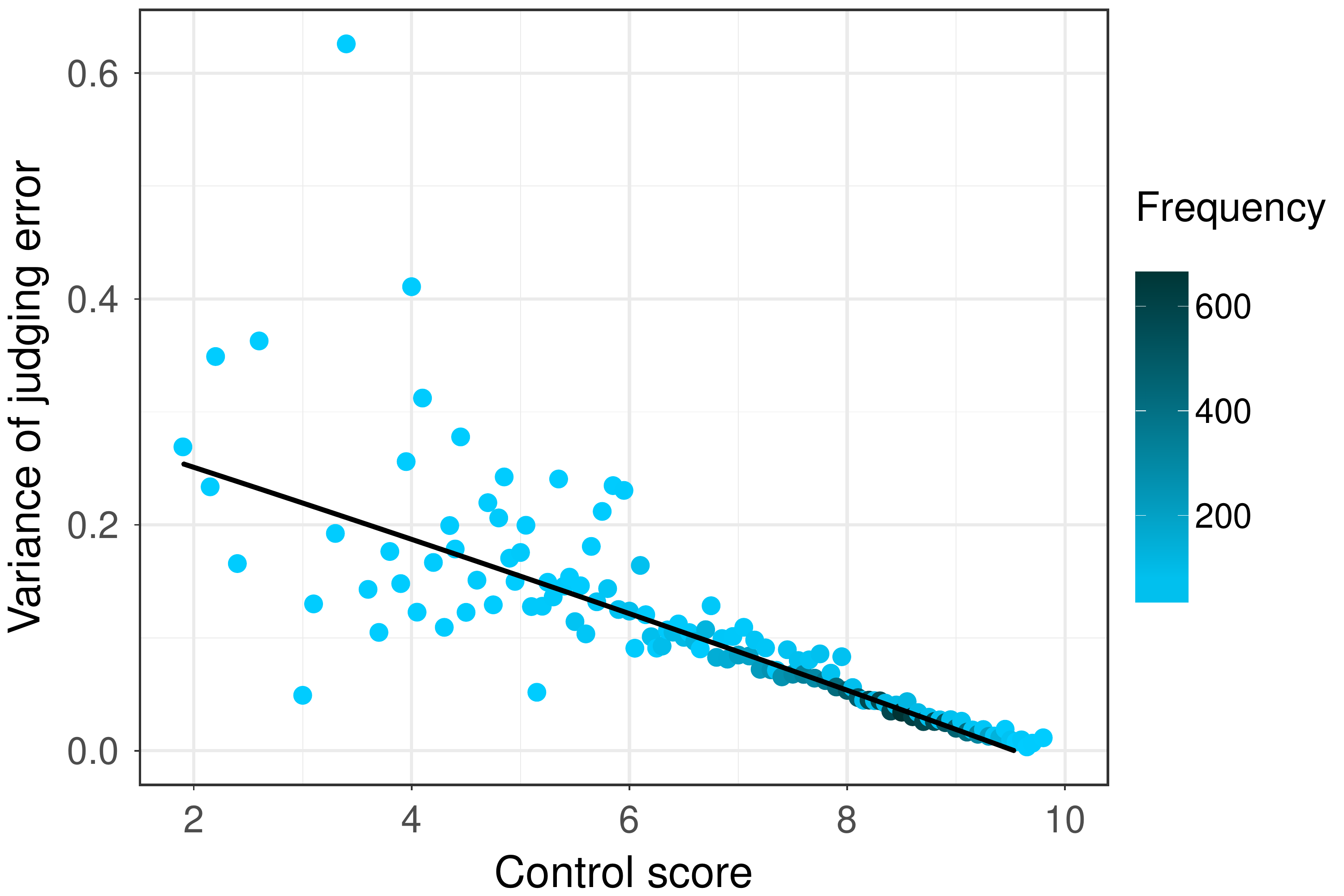}
	\caption{Variance of judging error versus control score in artistic gymnastics.}
	\label{fig:var:ag}
\end{figure}

\begin{figure}[h!]
	\centering
	\includegraphics[width=\columnwidth]{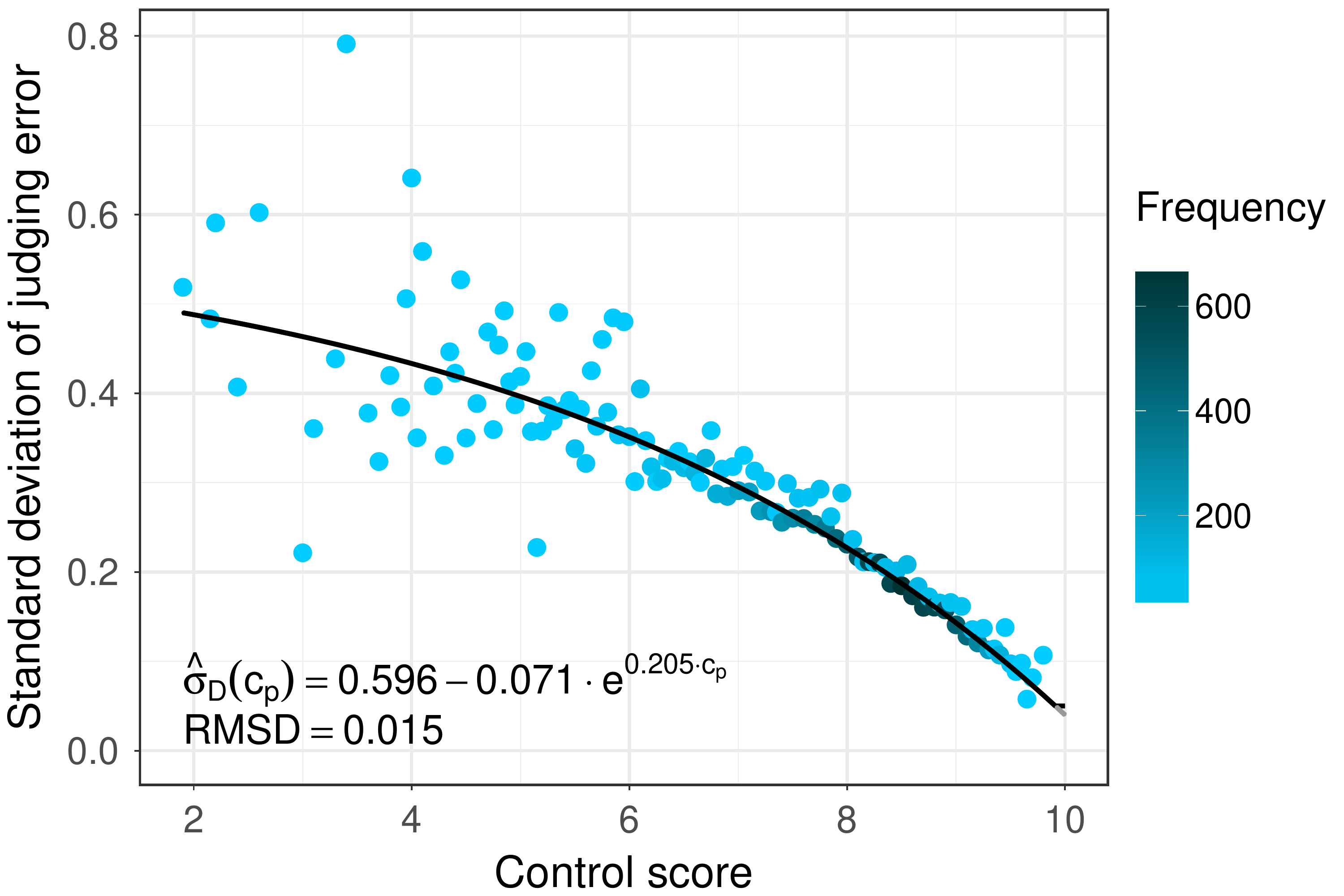}
	\caption{Standard deviation of judging error versus control score in artistic gymnastics, corresponding to the intrinsic judging error variability for this discipline.}
	\label{fig:sd:ag}
\end{figure}

Figures~\ref{fig:var:ag} and~\ref{fig:sd:ag} respectively show the sample variance and the sample standard deviation of the judging error $\hat{e}_{p,j}$ as a function of the control score $c_p$ for artistic gymnastics.
In Figures~\ref{fig:var:ag}, \ref{fig:sd:ag} and all similar figures that follow, the frequency is the number of performances with a given control score, and the fitted curves are exponential weighted least-squares regressions of the data. In Figure~\ref{fig:var:ag}, we observe that the sample variance decreases almost linearly with the control score, except for the best performances for which it does not converge to zero. By inspection, the fitted standard deviation in Figure~\ref{fig:sd:ag} is an outstanding fit. The outliers correspond to the rare gymnasts who  aborted or catastrophically missed their routine. The weighted root-mean-square deviation (RMSD) of the regression is 0.015, which is almost one order of magnitude smaller than the smallest deduction allowed by a judge. We use this exponential equation for our estimator of the standard deviation of the judging error $\hat{\sigma}_d(c_p)$, which we call the \emph{intrinsic judging error variability}.

\begin{figure}[H]
	\centering
	\includegraphics[width=\columnwidth]{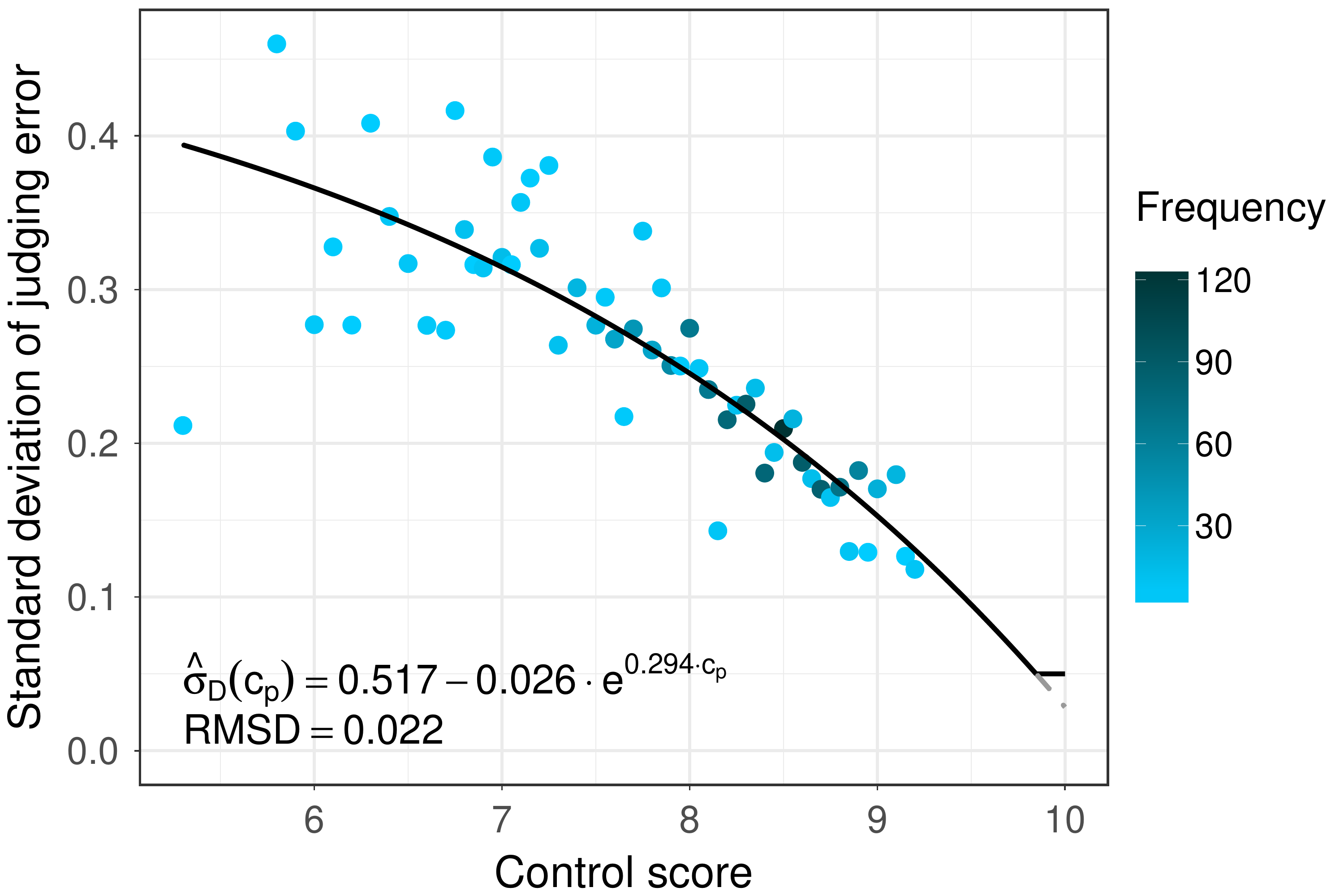}
	\caption{Standard deviation of judging error versus control score for still rings.}
	\label{fig:sd1}
\end{figure}

\begin{figure}[H]
	\centering
	\includegraphics[width=\columnwidth]{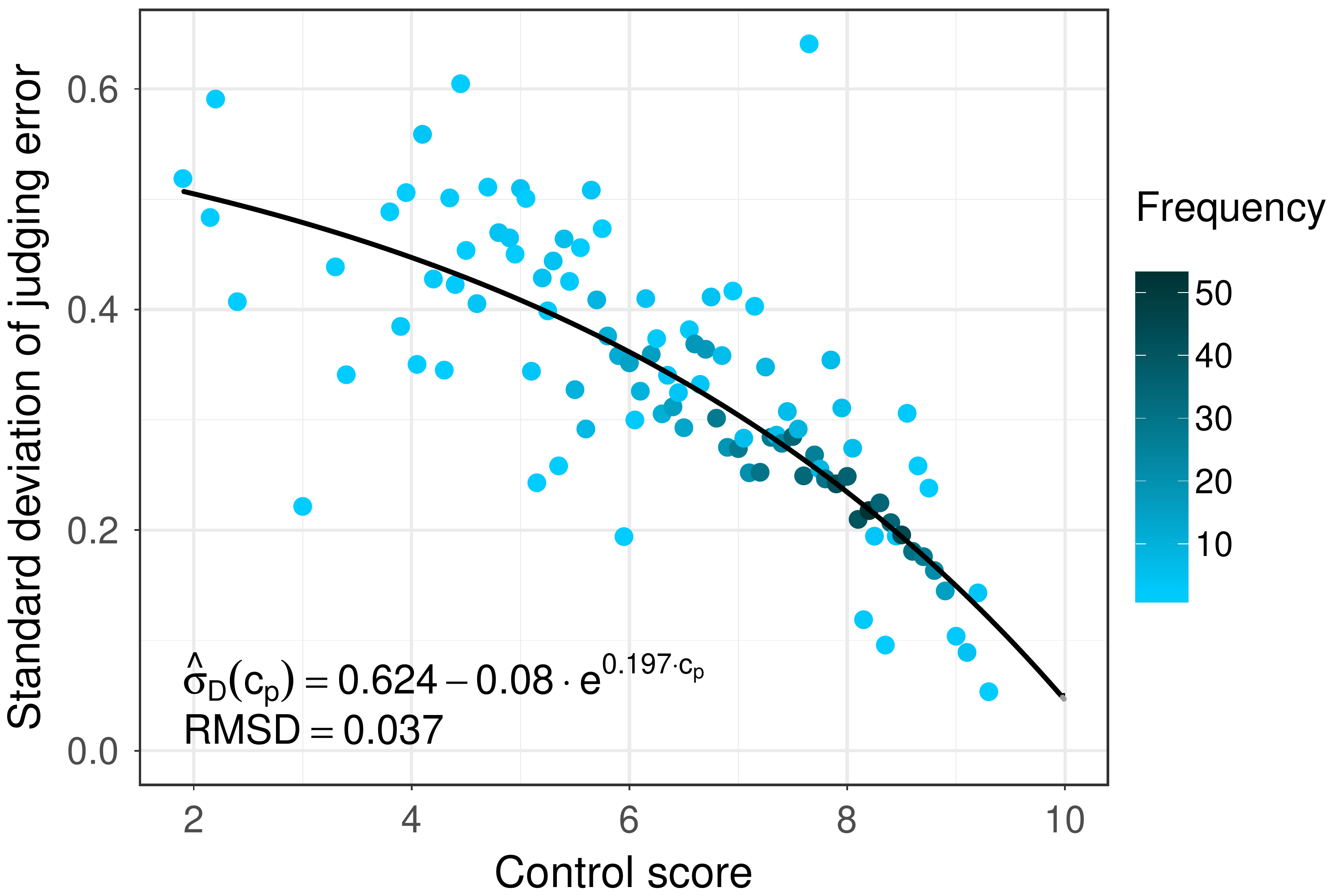}
	\caption{Standard deviation of judging error versus control score for uneven bars.}
	\label{fig:sd2}
\end{figure}

\begin{figure}[H]
	\centering
	\includegraphics[width=\columnwidth]{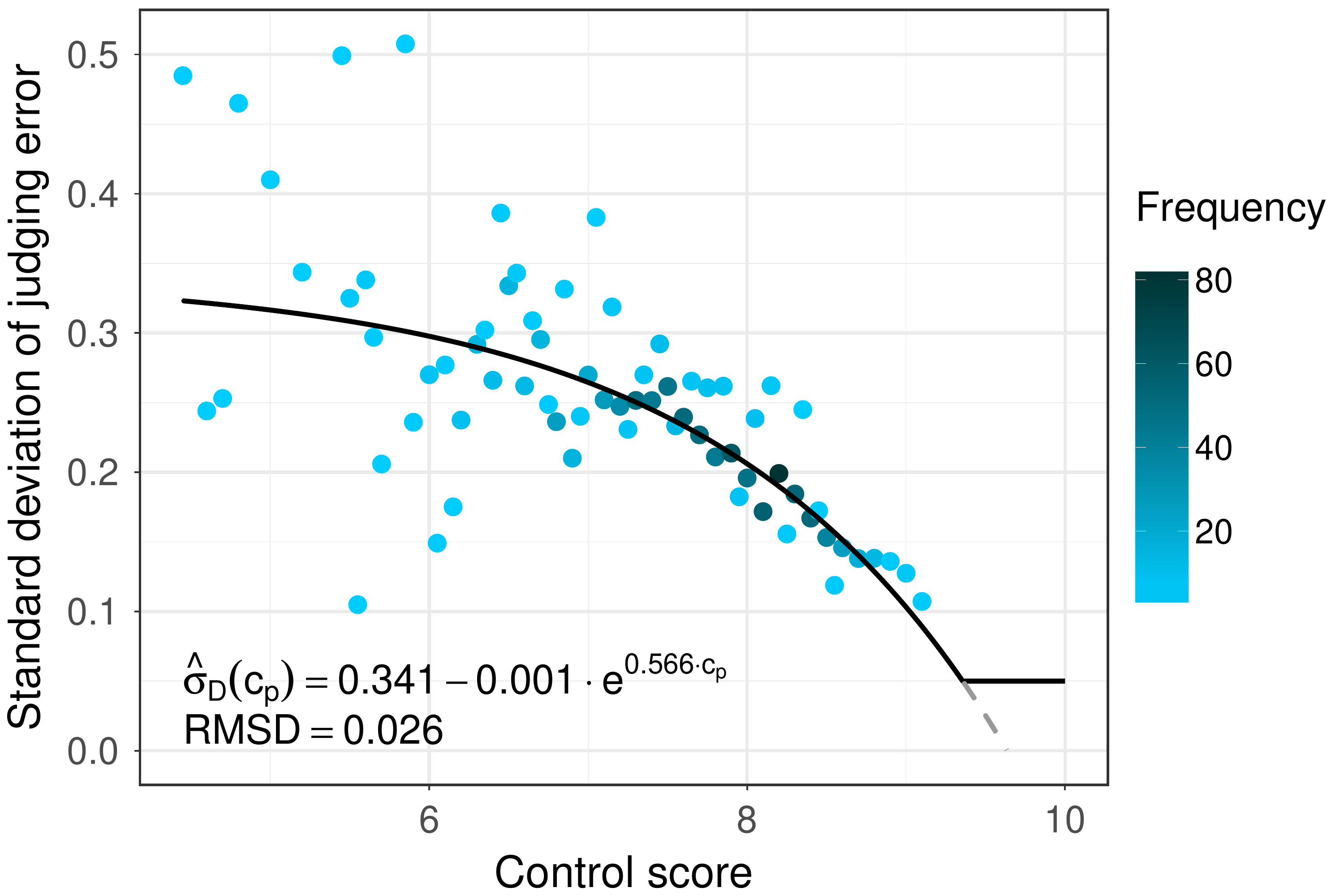}
	\caption{Standard deviation of judging error versus control score for women's floor exercise.}
	\label{fig:sd3}
\end{figure}

\begin{figure}[H]
	\centering
	\includegraphics[width=\columnwidth]{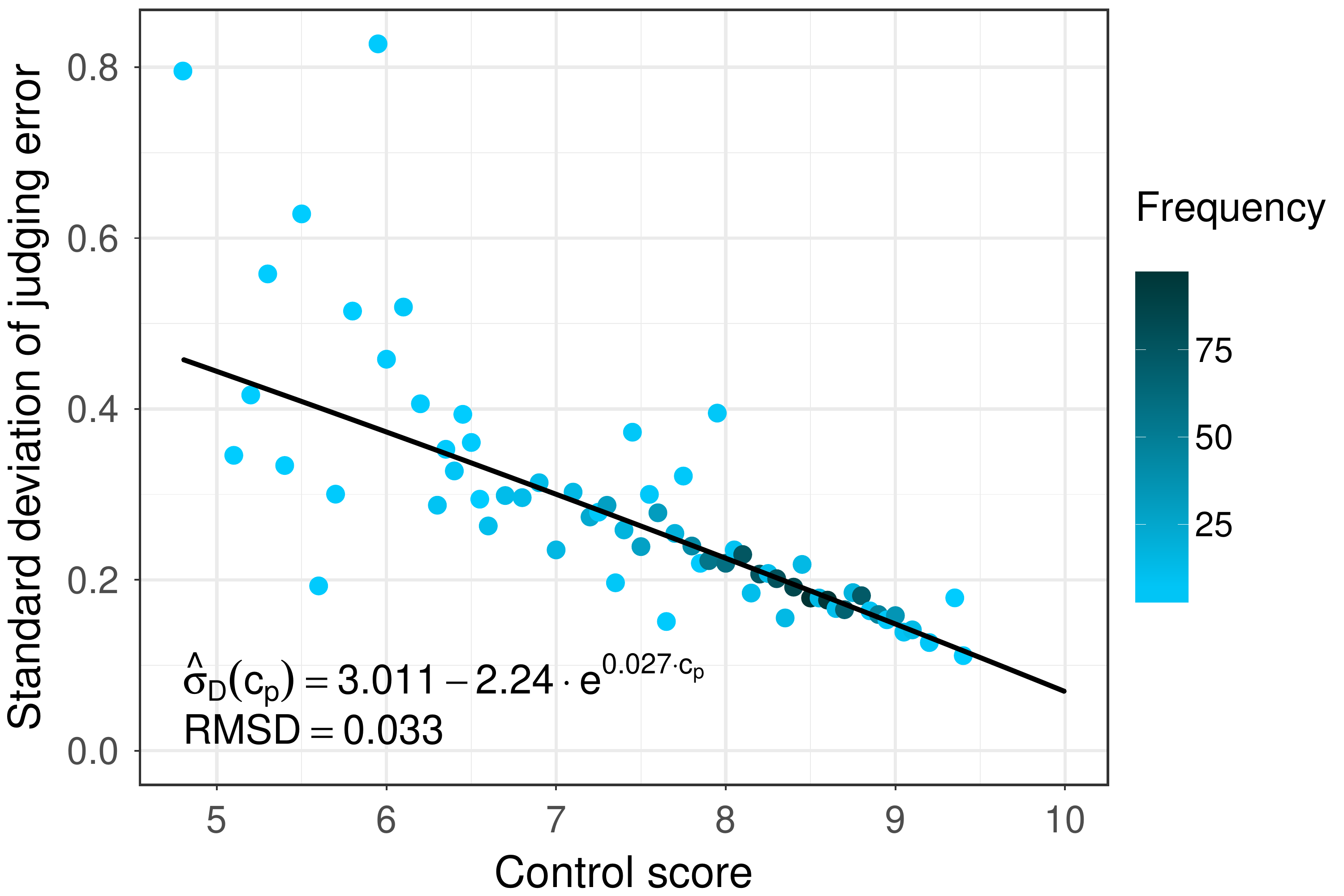}
	\caption{Standard deviation of judging error versus control score for men's floor exercise.}
	\label{fig:sd4}
\end{figure}

We can do the same analysis at the apparatus level. For example, Figures~\ref{fig:sd1}, \ref{fig:sd2}, \ref{fig:sd3} and \ref{fig:sd4} respectively show the intrinsic judging error variability (the weighted least-squares regression of the standard deviation of the judging error) for still rings, uneven bars, women's floor exercise and men's floor exercise.

More generally, the estimator for $\hat{\sigma}_d(c_p)$ depends on the discipline (or apparatus) $d$ under evaluation and the control score $c_p$ of the performance, and is given by
\begin{equation}
\label{eq:s1}
\hat{\sigma}_d(c_p) \triangleq \max(\alpha_d + \beta_d e^{\gamma_d c_p},0.05).
\end{equation}
For some apparatus like men's floor exercise in Figure~\ref{fig:sd4} the intrinsic judging error variability is linear within the data range. Since there is no mark close to 10 in our dataset, and since $\hat{\sigma}_d(c_p)$ becomes small for the best recorded performances, we can omit the mathematical ramifications of the bounded marking range. However, for apparatus such as women's floor exercise in Figure~\ref{fig:sd3}, the best fitted curves go to zero before 10. Since athletes might get higher marks than in our original data set in future competitions, we use $\max(\cdot,0.05)$ as a fail-safe mechanism to avoid comparing judges' marks to a very low and even negative extrapolated intrinsic error variability in the future.

We emphasize that all the disciplines and apparatus we analyzed have highly accurate regressions. Besides acrobatic gymnastics, for which we do not have as much data, the worst weighted root-mean-square deviation is $\text{RMSD} \approx 0.04$.

The marking score of Performance $p$ by Judge $j$ is
\begin{equation}
\label{eq:ms}
m_{p,j} \triangleq \frac{\hat{e}_{p,j}}{\hat{\sigma}_d(c_p)} = \frac{s_{p,j} - c_p}{\hat{\sigma}_d(c_p)}.
\end{equation}
It expresses the judging error as a function of the standard deviation for a specific discipline and control score. The overall marking score for Judge $j$ is given by 
\begin{equation}
\label{eq:4}
M_j \triangleq \sqrt{E[m_{p,j}^2]}=\sqrt{\frac{1}{n}\sum_{p=1}^n m_{p,j}^2}.
\end{equation}
The marking score of a perfect judge is 0, and a judge whose judging error is always equal to the intrinsic judging error variability $\hat{\sigma}_d(c_p)$ has a marking score of 1.0. The mean squared error weights outliers heavily, which is desirable for evaluating judges.

\begin{figure}[H]
	\centering
	\includegraphics[width=\columnwidth]{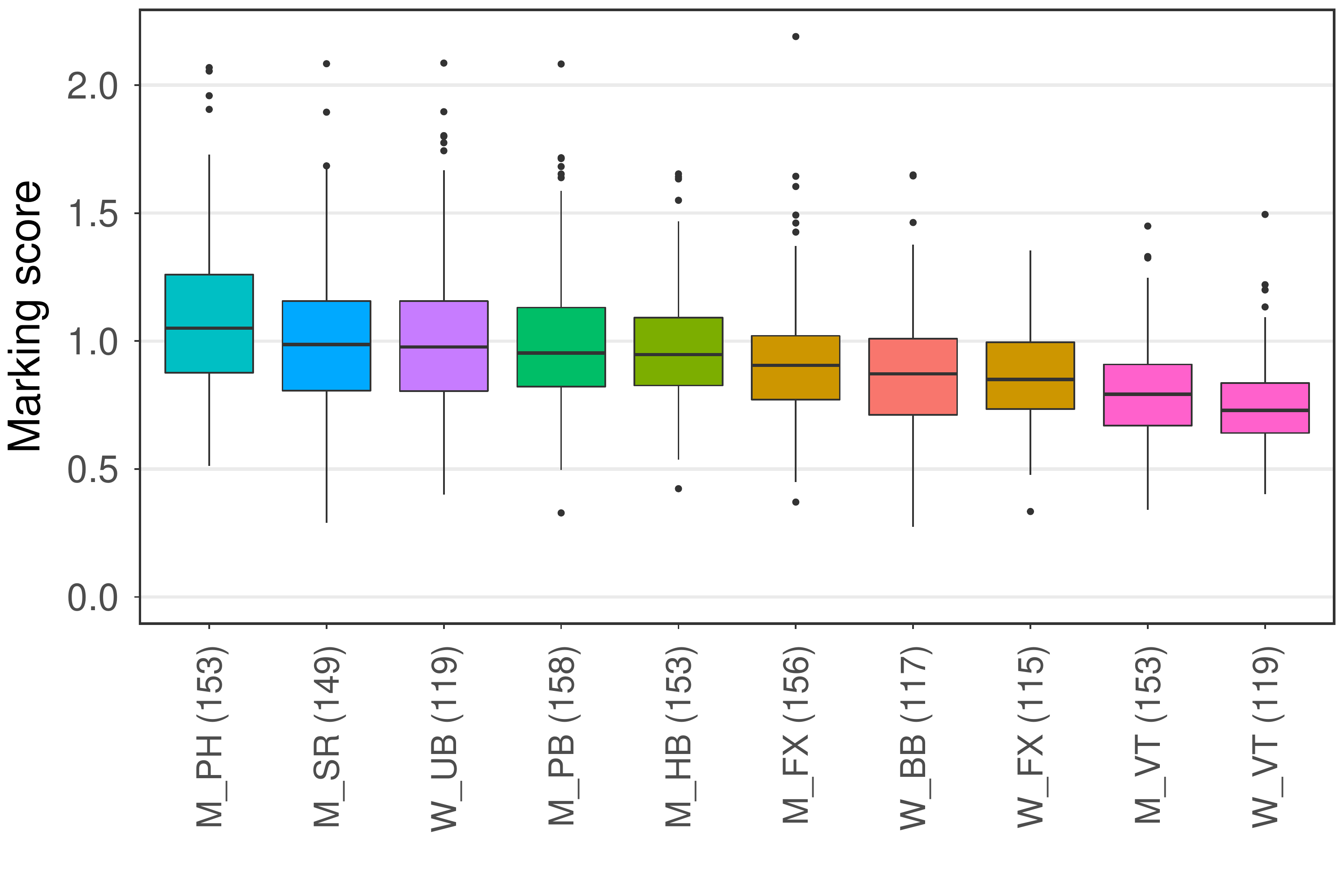}
	\caption{Distribution of the overall marking scores per artistic gymnastic apparatus using one overall formula. The acronyms are defined in Table~\ref{tab:apparatus}, and the numbers between brackets are the number of judges per apparatus in the dataset.}
	\label{fig:boxplots:ag:all}
\end{figure}

Figure~\ref{fig:boxplots:ag:all} shows the boxplots of the marking scores for all the judges for each apparatus in artistic gymnastics using the regression from Figure~\ref{fig:sd:ag}. The acronyms are defined in Table~\ref{tab:apparatus}. The first observation is that there are significant differences between apparatus. Pommel horse, for instance, is intrinsically more difficult to judge accurately than vault and floor exercise. The FIG confirms that the alternative, i.e., that judges in pommel horse are less competent than judges in men's vault or men's floor exercise, is highly unlikely. The differences between floor and vault on one side and pommel horse on the other side were previously observed in punctual competitions \cite{university-2009,london-2011,Bucar:2013}. Note that the better accuracy of vault judges does not make it easier to rank the gymnasts since many gymnasts execute the same jumps at a similar performance level. 

\begin{table}[H]
	\centering
	\begin{tabular}{ l l }
		\toprule
		Acronym & Apparatus \\
		\midrule
		BB 	& Balance beam (women)\\
		FX	& Floor exercise (men and women)\\
		HB 	& Horizontal bar (men)\\
		PB 	& Parallel bars (men) \\
		PH	& Pommel horse (men)\\
		SR 	& Still rings (men)\\
		UB 	& Uneven bars (women)\\
		VT 	& Vault (men and women)\\
		\bottomrule
	\end{tabular}
	\caption{The artistic gymnastics apparatus and their acronyms.}
	\label{tab:apparatus}
\end{table}

\begin{table}[H]
	\centering
	\begin{tabular}{ l l }
		\toprule
		Acronym & Apparatus \\
		\midrule
		DMT & Double mini-trampoline (men and women)\\
		IND	& Individual trampoline  (men and women)\\
		TUM & Tumbling (men and women)\\
		\bottomrule
	\end{tabular}
	\caption{The trampoline apparatus and their acronyms.}
	\label{tab:gt:apparatus}
\end{table}

\begin{figure}[H]
	\centering
	\includegraphics[width=\columnwidth]{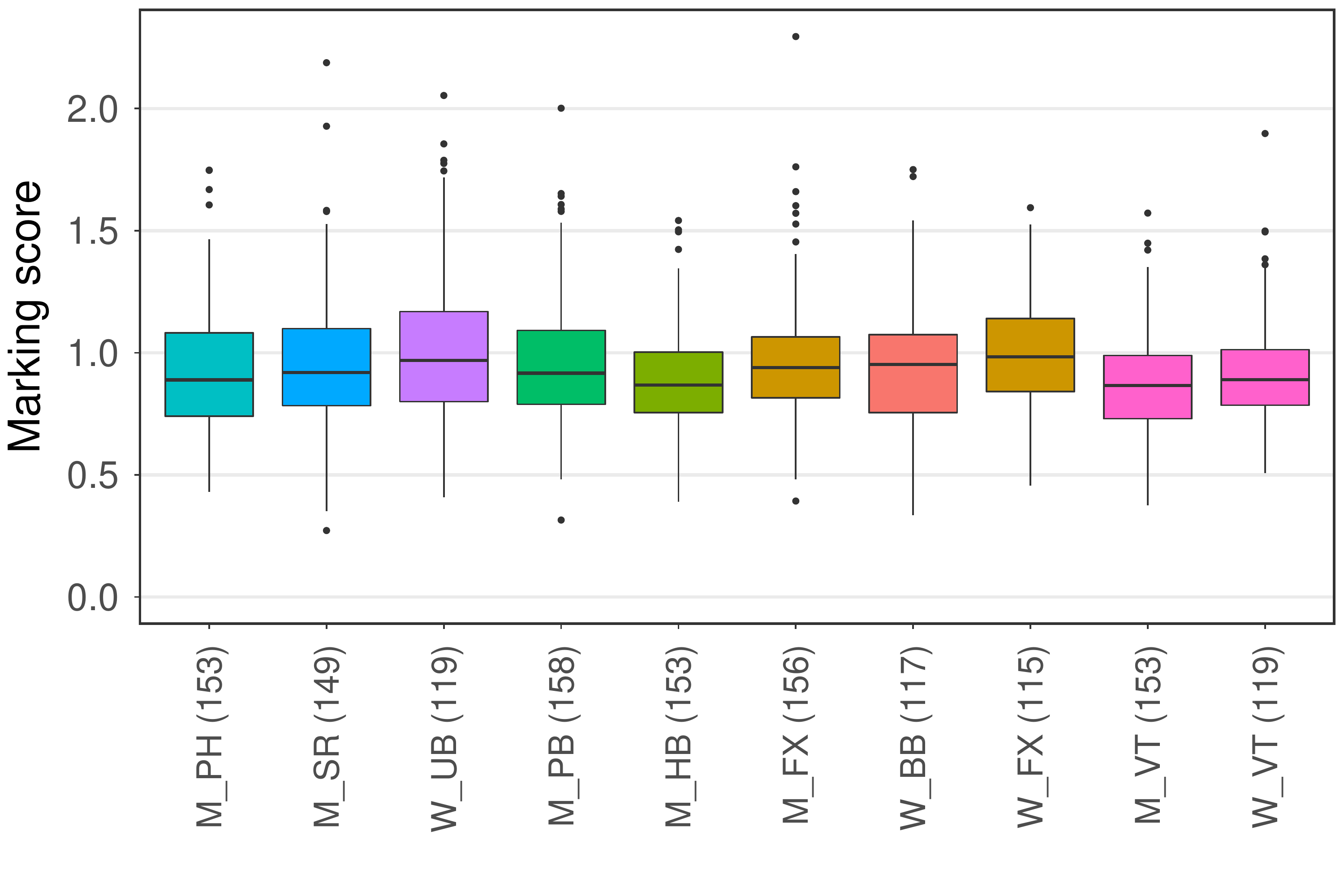}
	\caption{Distribution of the overall marking scores per artistic gymnastic apparatus using an individual formula per apparatus. The acronyms are defined in Table~\ref{tab:apparatus}, and the numbers between brackets are the number of judges per apparatus in the dataset.} 
	\label{fig:boxplots:ag:ind}
\end{figure}

A highly desirable feature for the marking score is to be comparable between apparatus and disciplines, which proves difficult with one overall formula. The differences between apparatus make it challenging for the FIG to qualitatively assess how good the judges are and to convey this information unambiguously to the interested parties. We thus estimated the intrinsic judging error variability $\hat{\sigma}_d(c_p)$ for each apparatus (instead of grouping them together) and used the resulting regressions to recalculate the marking scores. The results, presented in Figure~\ref{fig:boxplots:ag:ind}, now show a good uniformity and make it simpler to compare judges from different apparatus with each other. A pommel horse judge with a marking score of 1.0 is average, and so is a vault judge with the same marking score. This has allowed us to define a single set of quantitative to qualitative thresholds applicable across all the gymnastics apparatus and disciplines.

\subsection{Other gymnastic disciplines}
\label{subsection:other}

\begin{figure}[]
	\centering
	\includegraphics[width=\columnwidth]{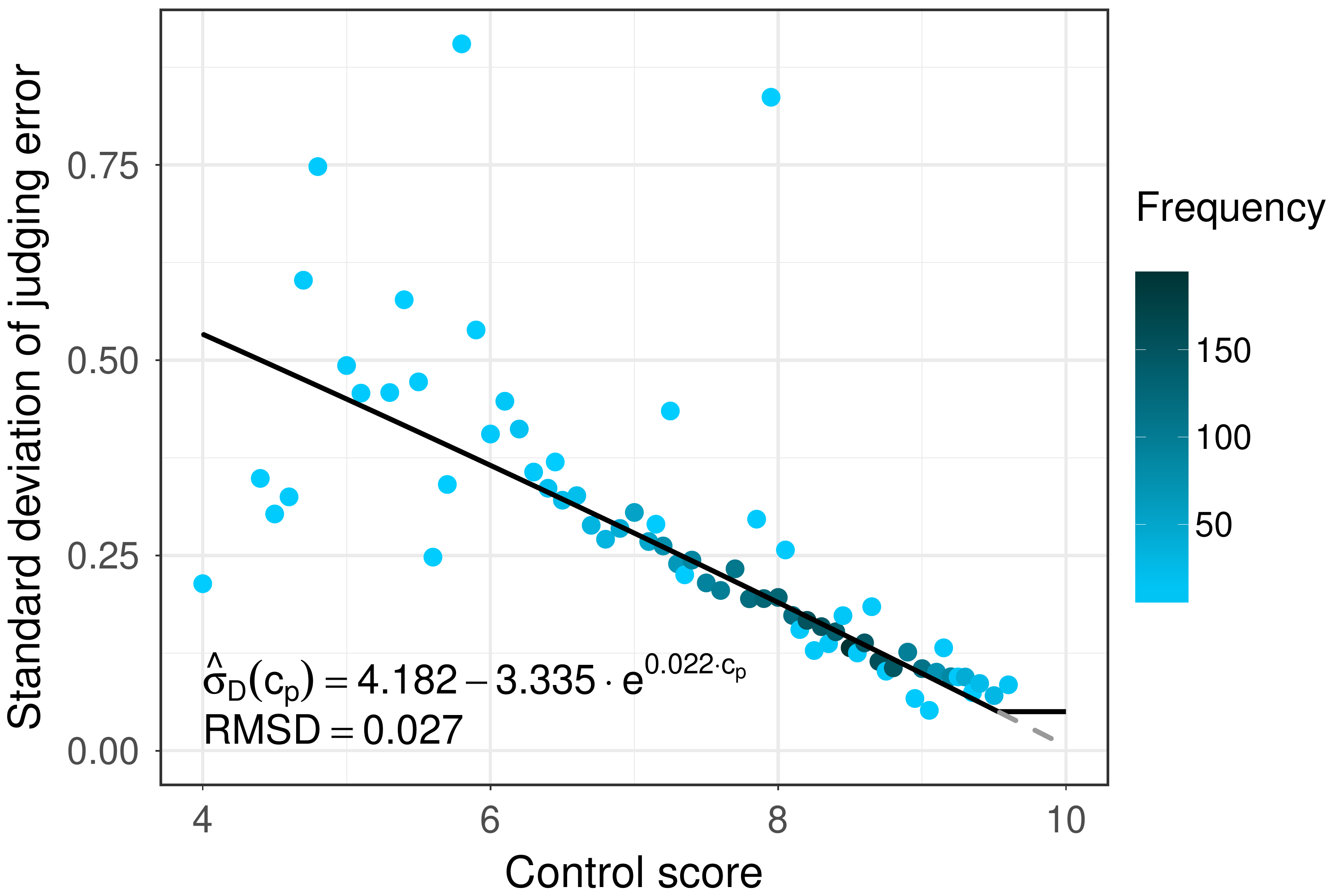}
	\caption{Standard deviation of judging error versus control score in rhythmic gymnastics.}
	\label{fig:sd:gr}
\end{figure}

\begin{figure}[]
	\centering
	\includegraphics[width=\columnwidth]{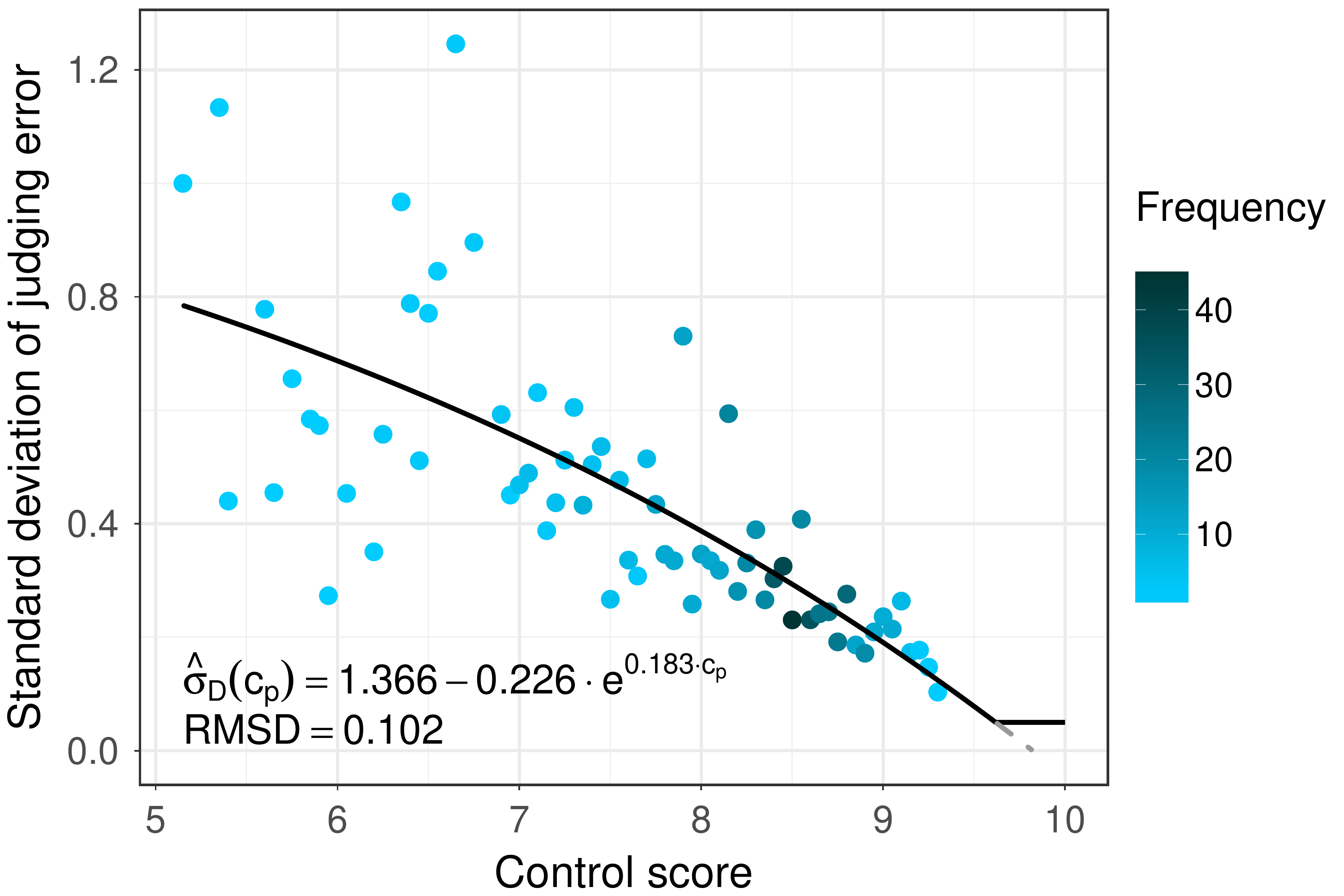}
	\caption{Standard deviation of judging error versus control score in acrobatic gymnastics.}
	\label{fig:sd:ac}
\end{figure}

\begin{figure}[]
	\centering
	\includegraphics[width=\columnwidth]{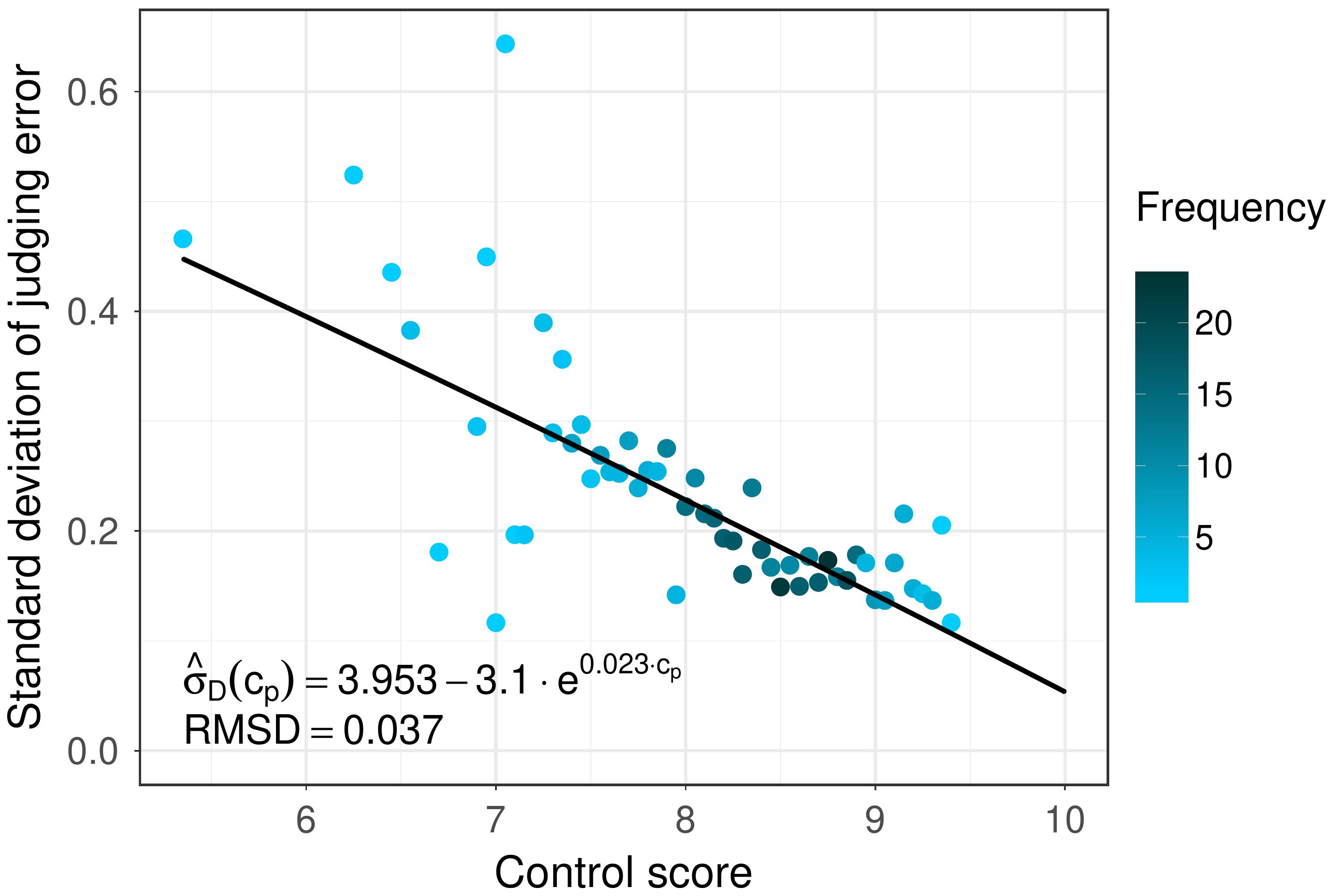}
	\caption{Standard deviation of judging error versus control score in aerobic gymnastics.}
	\label{fig:sd:ae}
\end{figure}

We use the same approach for the other gymnastics disciplines. Figures~\ref{fig:sd:gr}, \ref{fig:sd:ac} and \ref{fig:sd:ae} respectively show the weighted least-squares regressions for rhythmic gymnastics, acrobatic gymnastics and aerobic gymnastics. We do not discuss the results at the apparatus level, although we found notable differences: group routines in rhythmic gymnastics are more difficult to judge than individual ones, and groups in acrobatic gymnastics are more difficult to judge than pairs. We also analyzed the artistry judges in acrobatic and aerobic gymnastics, and were surprised to observe that the heteroscedasticity of their judging error was almost the same as for execution judges.

\begin{figure}[]
	\centering
	\includegraphics[width=\columnwidth]{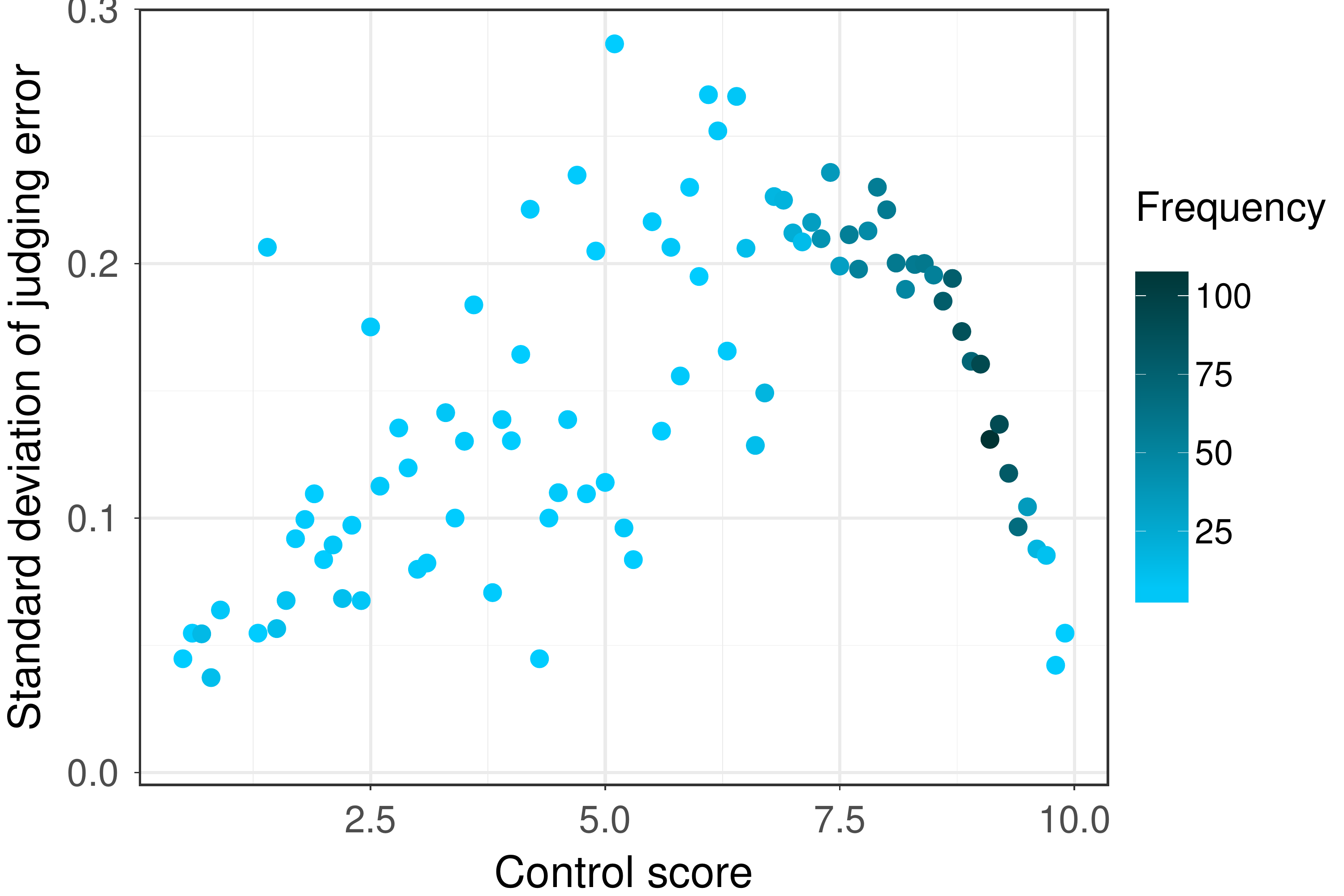}
	\caption{Standard deviation of judging error versus control score in trampoline.}
	\label{fig:sd:gt}
\end{figure}

\begin{figure}[]
	\centering
	\includegraphics[width=\columnwidth]{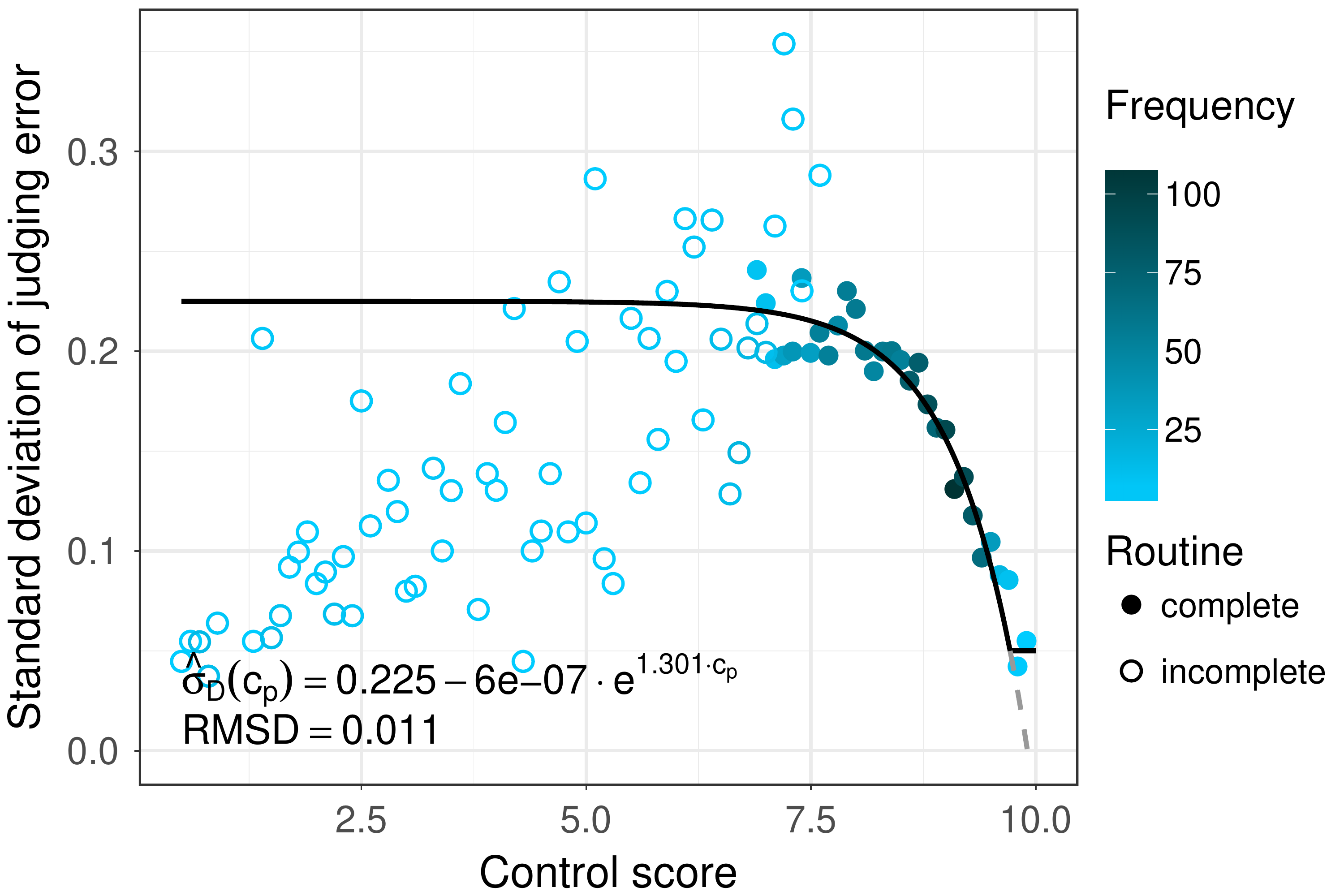}
	\caption{Standard deviation of judging error versus control score in trampoline. The rings indicate aborted routines. Data from synchronized trampoline is removed.}
	\label{fig:sd:gt:aborted}
\end{figure}

Trampoline, shown in Figure~\ref{fig:sd:gt}, was the most puzzling discipline to tackle. The behavior on the left side of the plot is due to gymnasts who aborted their routine before completing all their jumps, for instance by losing balance and landing a jump outside the center of the trampoline. We solved the problem by fitting the curves based on the completed routines. The result is shown in Figure~\ref{fig:sd:gt:aborted}, with aborted routines represented with rings instead of filled circles. Again, the weighted RMSD is excellent.

When calculating the marking score for trampoline judges, the marks of gymnasts who did not complete their exercise may be omitted. 
If they are accounted for, the estimator generously evaluates judges when gymnasts do not complete their routine, which results in a slightly improved overall marking score.

The behavior observed in trampoline appears in other sports with aborted routines or low scores~\cite{HM2018:heteroscedasticity} and can be modeled with concave parabola. This, however, decreases the accuracy of the regression for the best performances, which is undesirable. 

\begin{figure}[]
	\centering
	\includegraphics[width=\columnwidth]{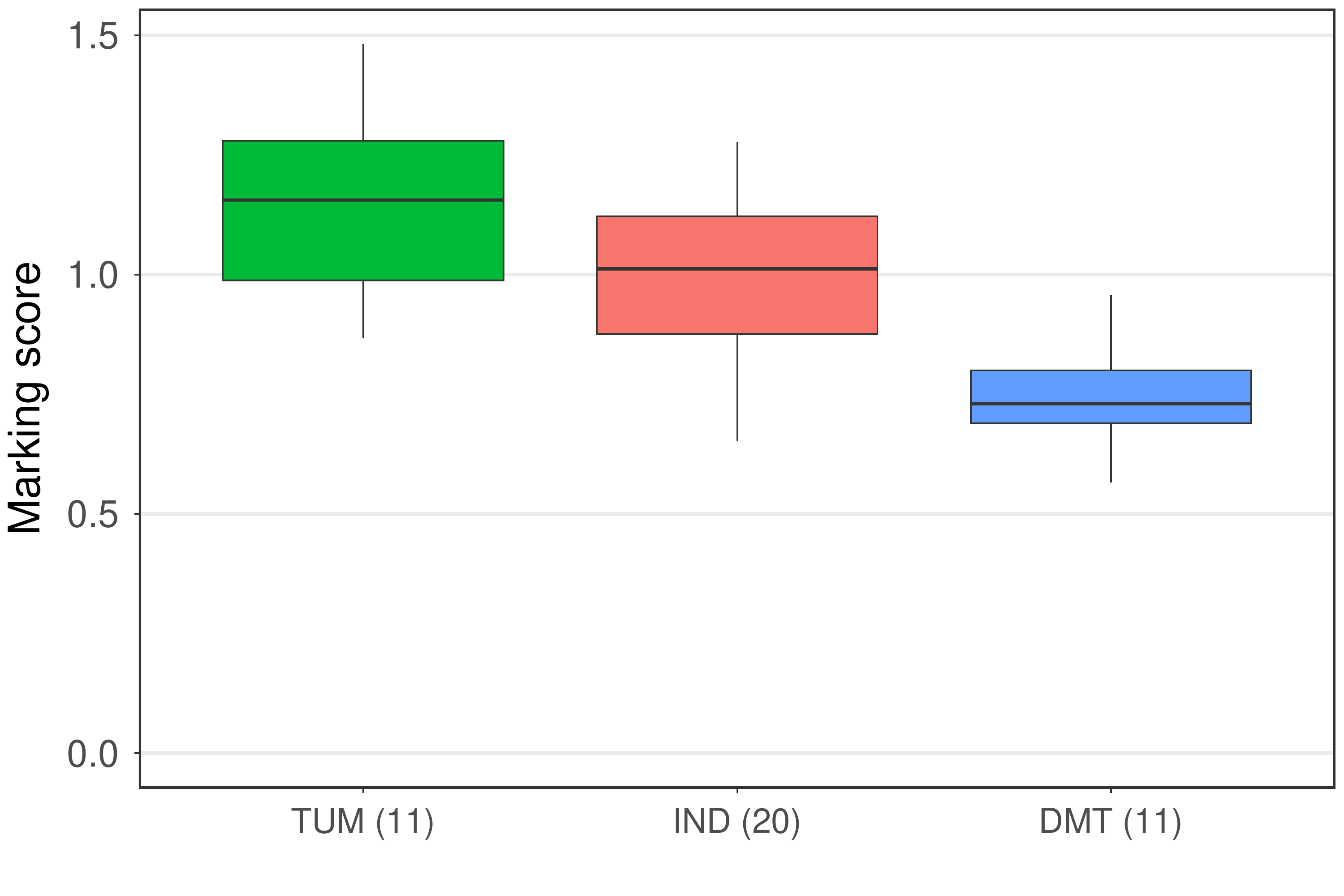}
	\caption{Distribution of the overall marking scores per trampoline apparatus using one overall formula. The acronyms are defined in Table~\ref{tab:gt:apparatus}, and the numbers between brackets are the number of judges per apparatus in the dataset.}
	\label{fig:boxplot:single:trampoline}
\end{figure}

Trampoline exhibits the largest differences between apparatus: tumbling is much more difficult to judge than individual trampoline, which in turn is much more difficult to judge than double mini-trampoline. The boxplots per trampoline apparatus in Figure~\ref{fig:boxplot:single:trampoline} clearly illustrate this (the acronyms are defined in Table~\ref{tab:gt:apparatus}). We thus use a different regression equation per apparatus. Finally, note that Figure~\ref{fig:sd:gt:aborted} excludes data from synchronized trampoline because its judging panels are partitioned in two halves, each monitoring a different gymnast. The subpanels (two judges each) are too small to derive accurate control scores.%

%% file: outlier.tex
\section{Outlier detection}
\label{sec:outlier}

\begin{figure}[h!]
	\vspace{-0mm}
	\centering
	\includegraphics[width=\columnwidth]{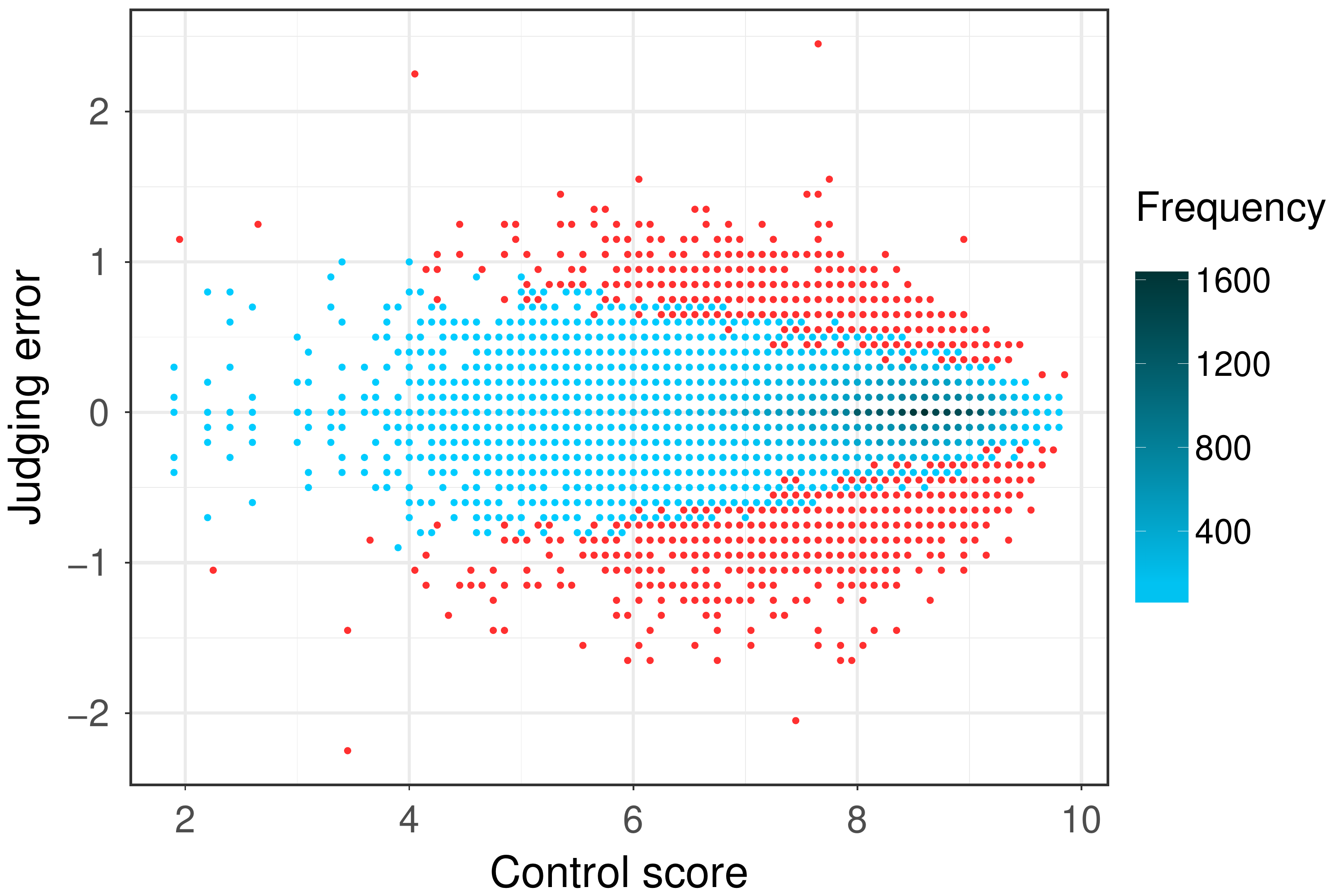}
	\caption{Distribution of the judging errors in artistic gymnastics. Dots in red are more than two standard deviations ($2 \cdot \sigma_d(c_a)$) away from the control score. To improve the visibility, we aggregate the points on a $0.1 \times 0.1$ grid and shift the outliers (red dots) by 0.05 on both axes.}  
	\label{fig:out1}
\end{figure}

We can use the marking score to signal judging marks that are unlikely high or low, with an increased emphasis on outliers from the same nationality. Figure~\ref{fig:out1}, like Figure~\ref{fig:D:ag}, shows the judging errors for artistic gymnastics judges. Differences of more than two standard deviations ($2 \cdot \hat{\sigma}_d(c_p)$) away from the control score are marked in red\footnote{We use a different equation for $\hat{\sigma}_d(c_p)$ per apparatus.}. The problem with this approach is that a bad judge has a lot of outliers, and a great judge none. This is not what the FIG wants, because an erratic judge can be unbiased and a precise judge can be dishonest.

\begin{figure}[h!]
	\centering
	\includegraphics[width=\columnwidth]{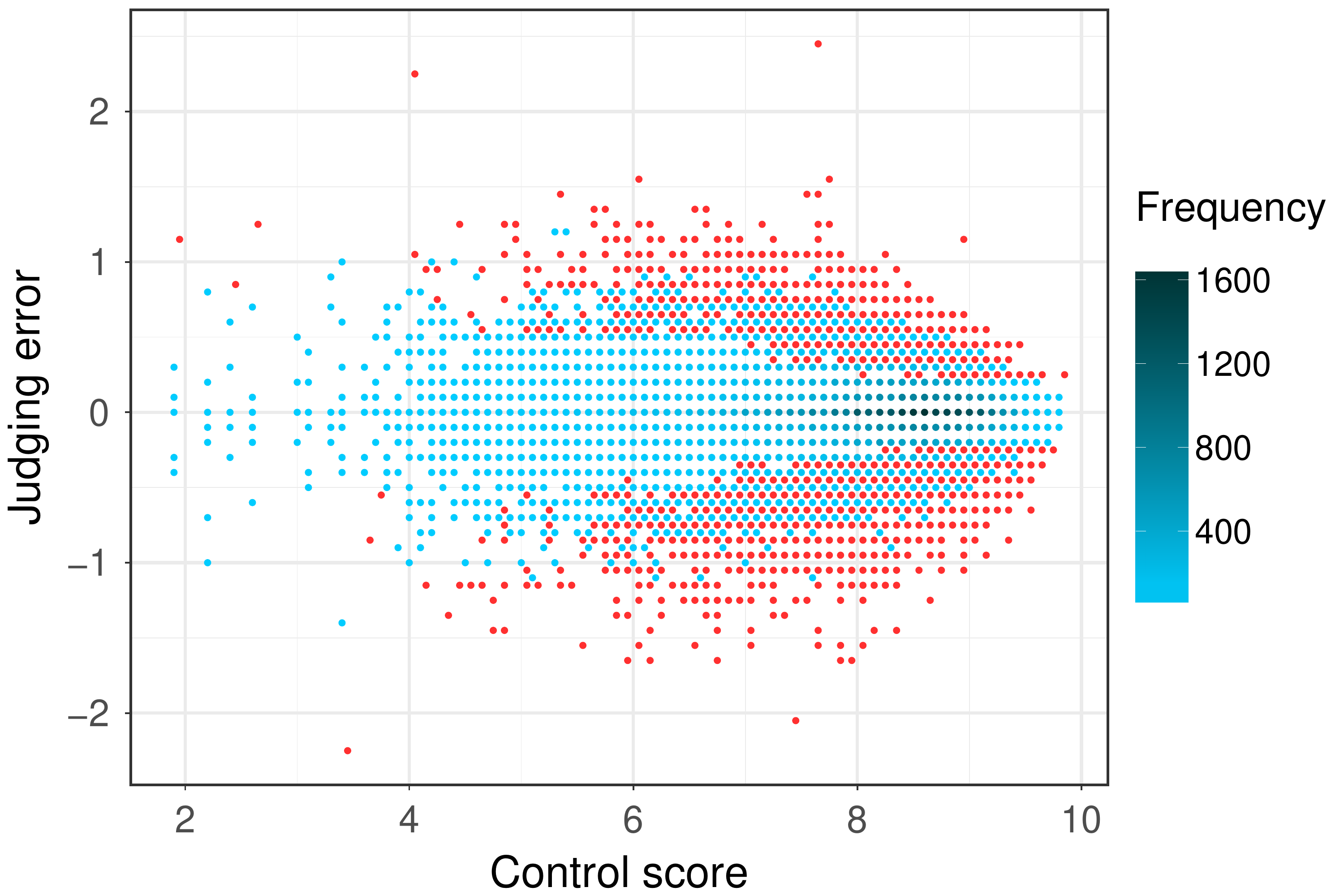}
	\caption{Distribution of the judging errors in artistic gymnastics. Dots in red are more than $2 \cdot \sigma_d(c_a) \cdot M_j$ away from the control score. To improve the visibility, we aggregate the points on a $0.1 \times 0.1$ grid and shift the outliers (red dots) by 0.05 on both axes.}
	\label{fig:out2}
	\vspace{-0mm}
\end{figure}

Instead of using the same standard deviation for all the judges, we scale the standard deviation by the overall marking score of each judge, and flag the judging scores that satisfy 
\begin{equation}
\label{eq:outlier}
|e_{p,j}| > \max(2\cdot\hat{\sigma}_d(c_p)\cdot M_j, 0.1).
\end{equation}
We use $\max(\cdot, 0.1)$ to ensure that a difference of 0.1 from the control score is never an outlier. The results are shown in Figure~\ref{fig:out2}. Eq.~\eqref{eq:outlier} flags $\approx 5\%$ of the marks, which is slightly more than what would be expected for a normal distribution. The advantage of the chosen approach is that it compares each judge to herself/himself, that is, it is more stringent for precise judges than for erratic judges. The disadvantage of the chosen approach is that one might think that a judge without outliers is good, which is false. The marking score and outlier detection work in tandem: a judge with a bad marking score is erratic, thus bad no matter how many outliers it has.

It is important to note that we cannot infer conscious bias, chicanery or cheating from an outlier mark. A flagged evaluation can be a bad but honest mistake, caused by external factors, or even indicate that a judge is out of consensus with the other judges who might be wrong at the same time. Nevertheless this information is useful for the FIG: performances with large discrepancies among panel judges systematically lead to careful video reviews post-competition. In egregious but very rare circumstances they may even result in sanctions by the FIG Disciplinary Commission. We present a comprehensive analysis of national bias in gymnastics in the second article of this series~\cite{HM2018:nationalbias}. 


%% file: ranking.tex
\section{Ranking score} \label{sec:rank}

The ranking of the gymnasts is determined by their scores, which are themselves aggregated from the marks given by the judges. The old iteration of JEP used a rudimentary \emph{ranking score} to evaluate to what extent judges ranked the best athletes in the right order. In a vacuum this makes sense: the FIG wants to select the most deserving gymnasts for the finals, and award the medals in the correct order. In this section we show that providing an objective assessment of the judges based on the order in which they rank the best athletes is problematic, and we recommended that the FIG stops using this approach.

\begin{defn}
	Let $G = \{g_1, g_2, \dots, g_n\}$ be a set of $n$ gymnasts. A \emph{ranking} on $G$ is a sequence $r = a_1a_2a_3 \dots a_n$, $a_i \neq a_j$ $\forall i, j \in \{1, \dots, n\}$ of all the elements of $G$ that defines a weak order on $G$. Alternatively, a ranking can be noted as $r = (r_{g_1}, r_{g_2}, r_{g_3}, \dots )$, where $r_{g_1}$ is the rank of Gymnast $g_1$, $r_{g_2}$ is the rank of Gymnast $g_2$, and so on.
\end{defn}

The mathematical comparison of rankings is closely related to the analysis of voting systems and has a long and rich history dating back to the work of Ramon Llull in the 13th century. Two popular metrics on the set of weak orders are Kendall’s $\tau$ distance \cite{Ken1938} and Spearman’s footrule \cite{Spe1904}, both of which are within a constant fraction of each other \cite{Diaconis:1977}. In recent years, \textcite{Kumar:2010} generalized these two metrics by taking into account element weights, position weights, and element similarities. Their motivation was to find the ranking minimizing the distance to a set of search results from different search engines.

\begin{defn}
	Let $r$ be a ranking of $n$ competitors. Let $w = (w_1, \dots, w_n)$ be a vector of element weights. Let $\delta = (\delta_1, \dots, \delta_n)$ be a vector of position swap costs where $\delta_1 \triangleq 1$ and $\delta_i$ is the cost of swapping elements at positions $i-1$ and $i$ for $i\in\{2,3,\dots,n\}$. Let $p_i = \sum\limits_{j=1}^{i} \delta_j$ for $i\in\{1,2,\dots,n\}$. We define the mean cost of interchanging positions $i$ and $r_i$ by $\bar{p}(i) = \frac{p_i - p_{r_i}}{i - r_i}$. Finally, let $D:\{1, \dots, n\} \times \{1, \dots, n\}$ be a non-empty metric and interpret $D(i, j) = D_{ij}$ as the cost of swapping elements $i$ and $j$. The generalized Kendall’s $\tau$ distance \cite{Kumar:2010} is
	\begin{equation}
	\label{eq:gkt}
	K’^{\ast} = K’_{w,\delta,D}(r) = \sum_{s > t} w_s w_t \bar{p}_s \bar{p}_t D_{st} [r_s < r_t].
	\end{equation}
\end{defn}

Note that $K^{'*}$ is the distance between $r$ and the identity ranking $id = (1, 2, 3, \dots)$.  To calculate the distance between two rankings $r^1$ and $r^2$, we calculate $K’(r^1, r^2) = K’_{w,\delta,D}(r^1 \circ (r^2)^{-1})$, where $(r^2)^{-1}$ is the right inverse of $r^2$.

These generalizations are natural for evaluating gymnastics judges: swapping the gold and silver medalists should be evaluated more harshly than inverting the ninth and tenth best gymnasts, but swapping the gold and silver medalists when their marks are 9.7 and 9.6 should be evaluated more leniently than if their marks are 9.7 and 8.7.

\begin{table}
	\centering
	\begin{tabular}{ c  | ccc }
		\toprule
		Parameter set 	& $w_i$	& $\delta_i$	& $D_{ij}$ 	\\
		\midrule
		1			& 1	& 1		& 1 \\
		2			& 1	& 1 & $|c_i-c_j|$ \\  
		3			& 1	& $\frac{1}{i}$ & $|c_i-c_j|$ \\
		\bottomrule
	\end{tabular}
	\caption{Parameters of the ranking scores for our two series of simulations.}
	\label{tab:rs}
\end{table}

\begin{figure}[h!]
	\centering
	\includegraphics[width=\columnwidth]{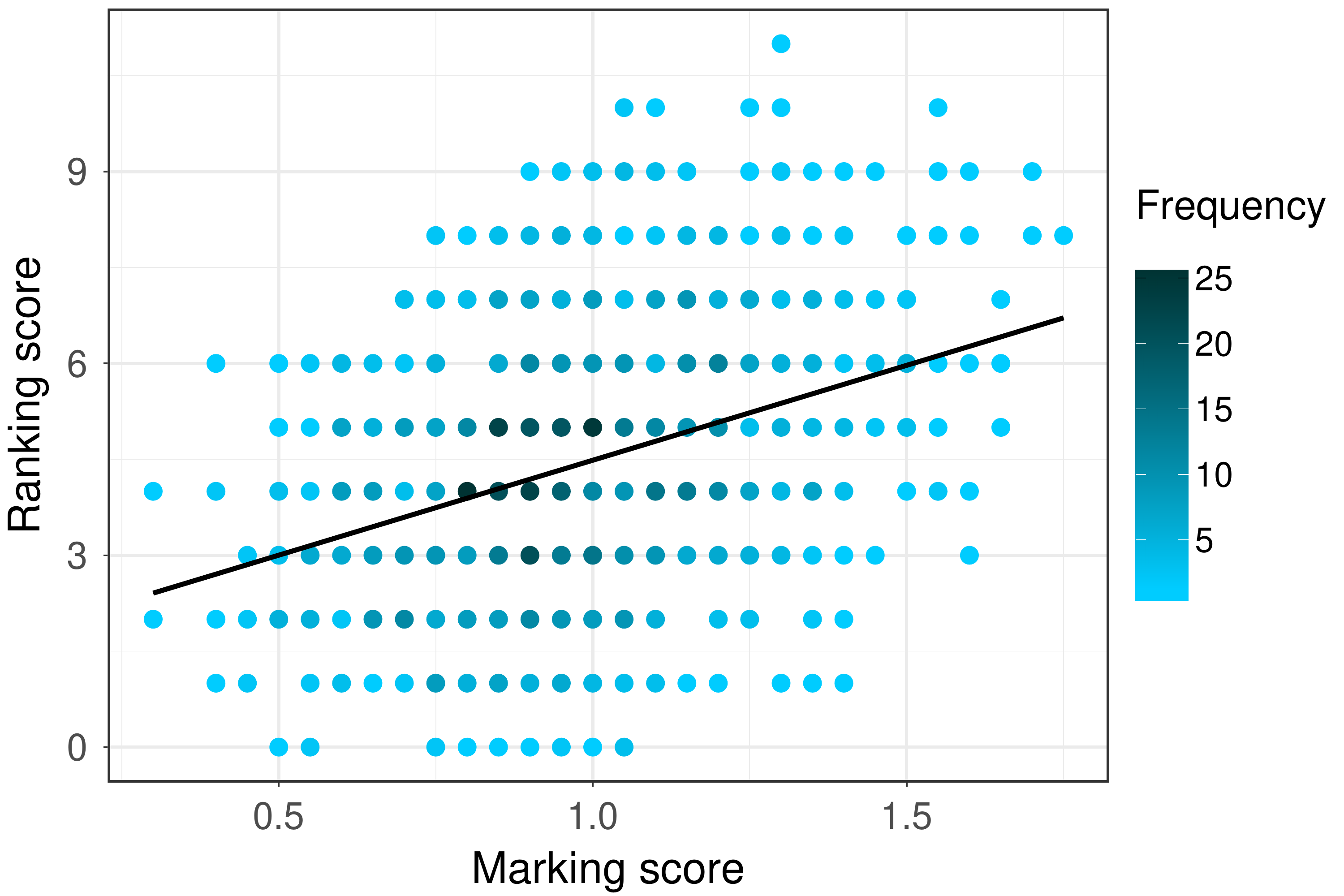}
	\caption{Ranking score vs marking score for 1000 synthetic average judges and the first set of ranking score parameters from Table~\ref{tab:rs}. We aggregate the points on the x-axis to improve visibility.}
	\vspace{-0mm}
	\label{fig:sim1}
\end{figure}

\begin{figure}[h!]
	\centering
	\includegraphics[width=\columnwidth]{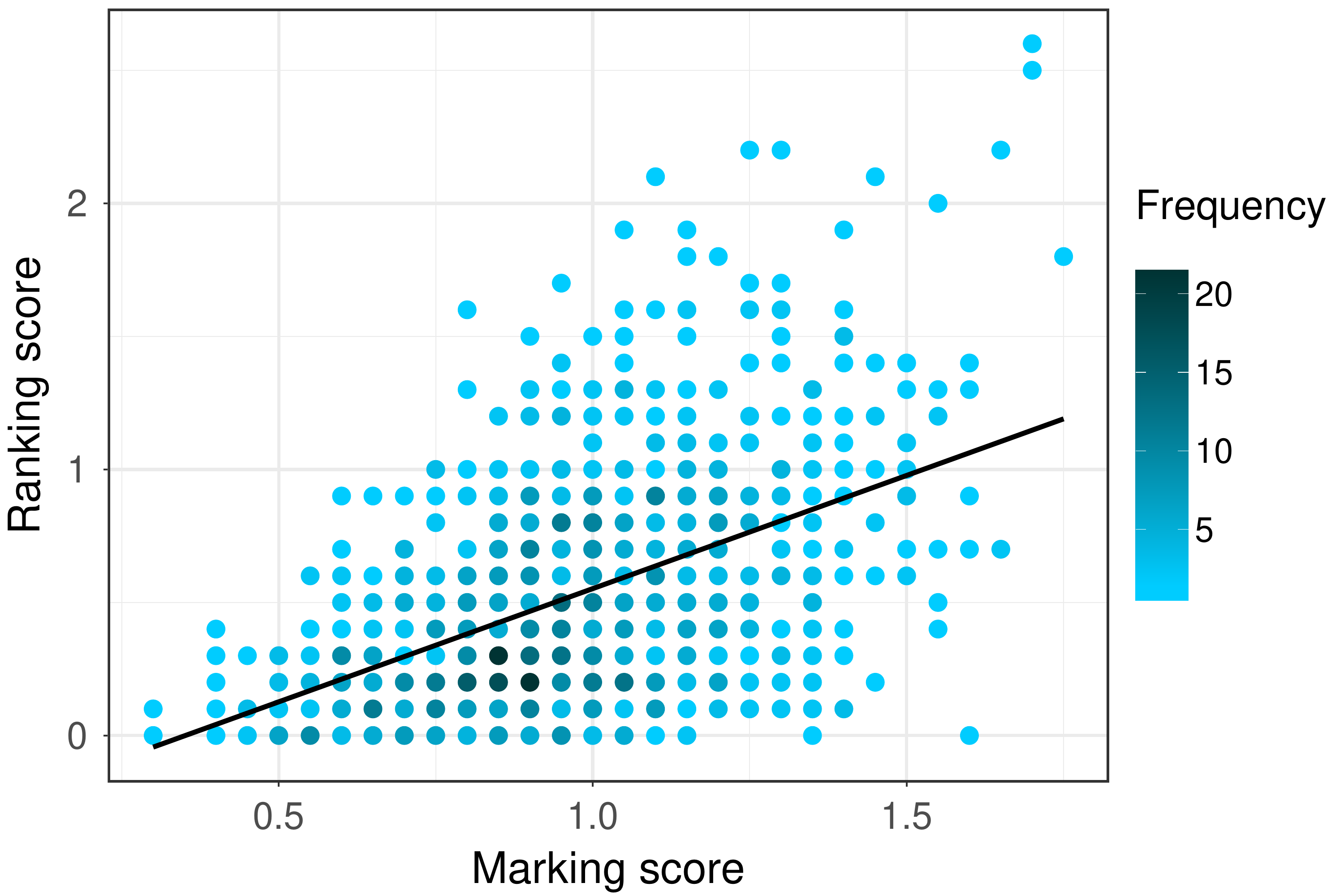}    \caption{Ranking score vs marking score for 1000 synthetic average judges and the second set of ranking score parameters from Table~\ref{tab:rs}. We aggregate the points to improve visibility.}
	\label{fig:sim2}
\end{figure}

\begin{figure}[h!]
	\centering
	\includegraphics[width=\columnwidth]{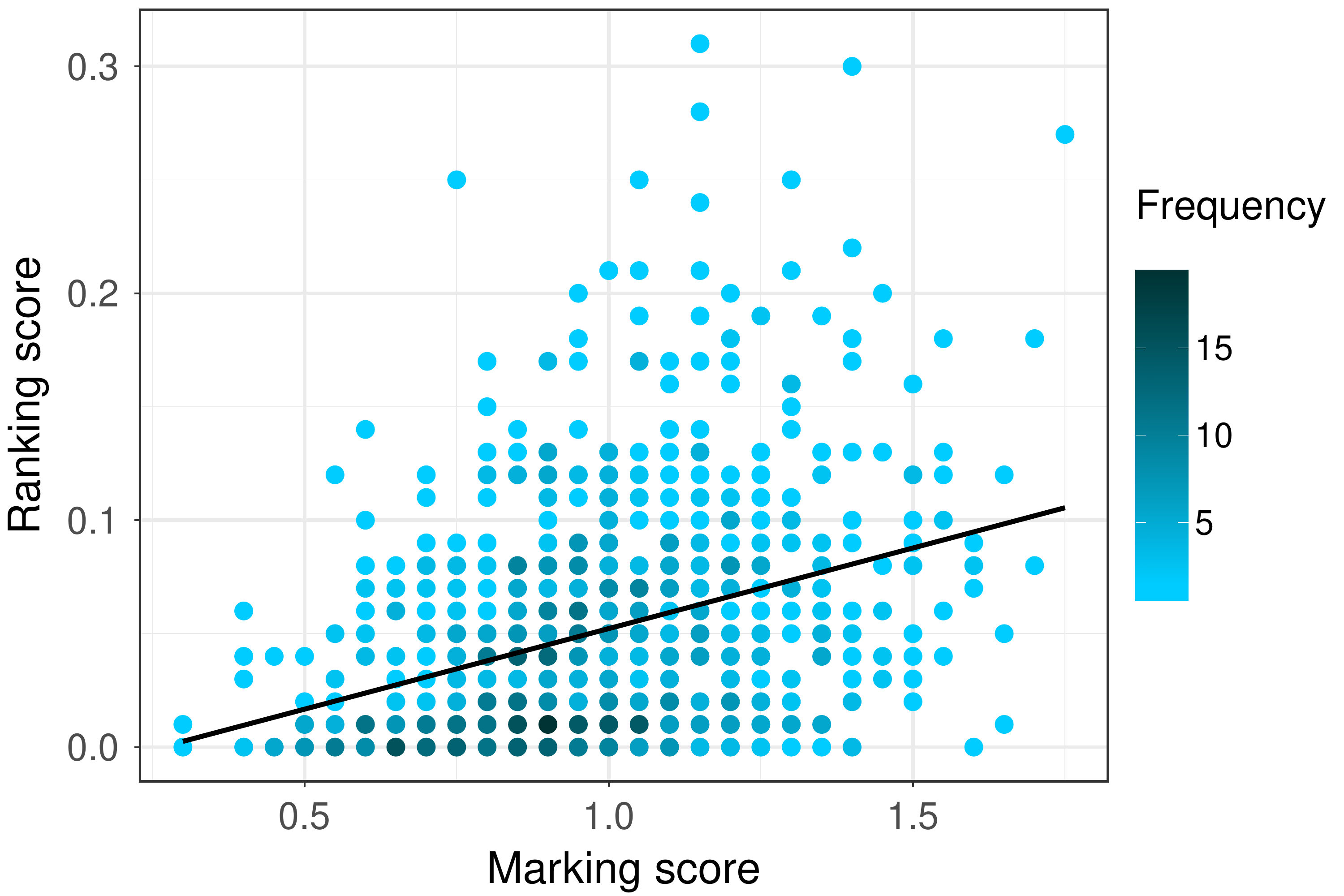}    \caption{Ranking score vs marking score for 1000 synthetic average judges and the third set of ranking score parameters from Table~\ref{tab:rs}. We aggregate the points to improve visibility.}
	\label{fig:sim3}
\end{figure}

To test the relevance of ranking scores as a measurement of judging accuracy, we ran several simulations to compare them to our marking score. As an example for this article, we use the men's floor exercise finals at the 2016 Rio Olympic Games. We first calculate the control scores $c_1, c_2, \dots, c_8$ of the eight finalists from the marks given by the seven execution judges (five panel judges and two reference judges). We then simulate the performance of 1000 average judges $j\in\{1,2,...,1000\}$ by randomly creating, for each of them, eight marks $s_{1,j}, s_{2,j}, \dots, s_{8,j}$ for the eight finalists using a normal distribution with mean $c_p$ and standard deviation $\hat{\sigma}_d(c_p)$ for $p\in\{1,2,\dots,8\}$. We then calculate, for each judge, the marking score as well as three ranking scores based on Eq.~(\ref{eq:gkt}) with the three different sets of parameters from Table~\ref{tab:rs}.

Figures~\ref{fig:sim1}, \ref{fig:sim2} and \ref{fig:sim3} show the ranking score with respect to the marking score of the 1000 judges for the three parameter sets. The figures illustrate that the correlation between the ranking score and the marking score varies widely depending on the chosen parameters.

The parameters used in Figure~\ref{fig:sim1} are those of the original version of Kendall’s $\tau$ distance \cite{Ken1938}. This simply counts the number of bubble sort swaps required to transform one ranking into the other; swapping the first and second gymnasts separated by 0.1 point is equivalent to swapping the seventh and eighth gymnasts separated by 1.0 point. In Figure~\ref{fig:sim2}, the element swap costs vary $(D_{ij}=|c_i-c_j|)$. This decreases the penalty of swaps as the marks get closer to each other; in particular, swapping two gymnasts with the same control score $c_i=c_j$ incurs no penalty.
This increases the correlation between the marking score and the ranking score, and both, to some extent, measure the same thing. In Figure~\ref{fig:sim3}, we also vary the position swap costs $(\delta_i=\frac{1}{i})$. This increases the importance of having the correct order as we move towards the gold medalist. The correlation between the marking score and the ranking score decreases, thus we penalize good judges that unluckily make mistakes at the wrong place, and reward erratic judges that somehow get the podium in the correct order.

It is unclear how to parametrize the ranking score; it is either redundant with the marking score, or too uncorrelated to be of any practical value. The {marking score} already achieves our objectives. It is based on the theoretical performances of the gymnasts over hundreds of performances for each judge and reflects bias and cheating, as these involve changing the marks up or down for some of the performances. Furthermore, the FIG is adamant that a theoretical judge who ranks all the gymnasts in the correct order but is either always too generous or too strict is not a good judge because he/she does not apply the Codes of Points properly. From these observations,  the FIG stopped using ranking scores to monitor the accuracy of its judges.

%% file: observations.tex
\section{Observations, discoveries and recommendations} \label{sec:obs}

During the course of this work we made interesting and sometimes surprising observations and discoveries that led to recommendations to the FIG. We summarize our observations about reference  judges in Section~\ref{subsection:reference} and judging gender discrepancies in Section~\ref{subsection:gender}.

\subsection{Reference judges}

\label{subsection:reference}

In addition the regular panel of execution judges, all the gymnastic disciplines except trampoline also have so called \emph{reference judges}. In artistic and rhythmic gymnastics, there are two reference judges, and the aggregation process is as follows\footnote{Acrobatic and aerobic gymnastics have a similar process for execution and artistry judges.}. The execution panel score is the trimmed mean of the middle three of out five execution panel judges, and the reference score is the arithmetic mean of the two reference judges' marks.
If the gap between the execution panel score and the reference score exceeds a predefined tolerance threshold, and if the difference between the marks of both reference judges is below a second threshold, then the final execution score of the gymnast is the mean of the execution panel and reference scores. This makes reference judges dangerously powerful. 

\begin{figure}[H]
	\centering
	\includegraphics[width=\columnwidth]{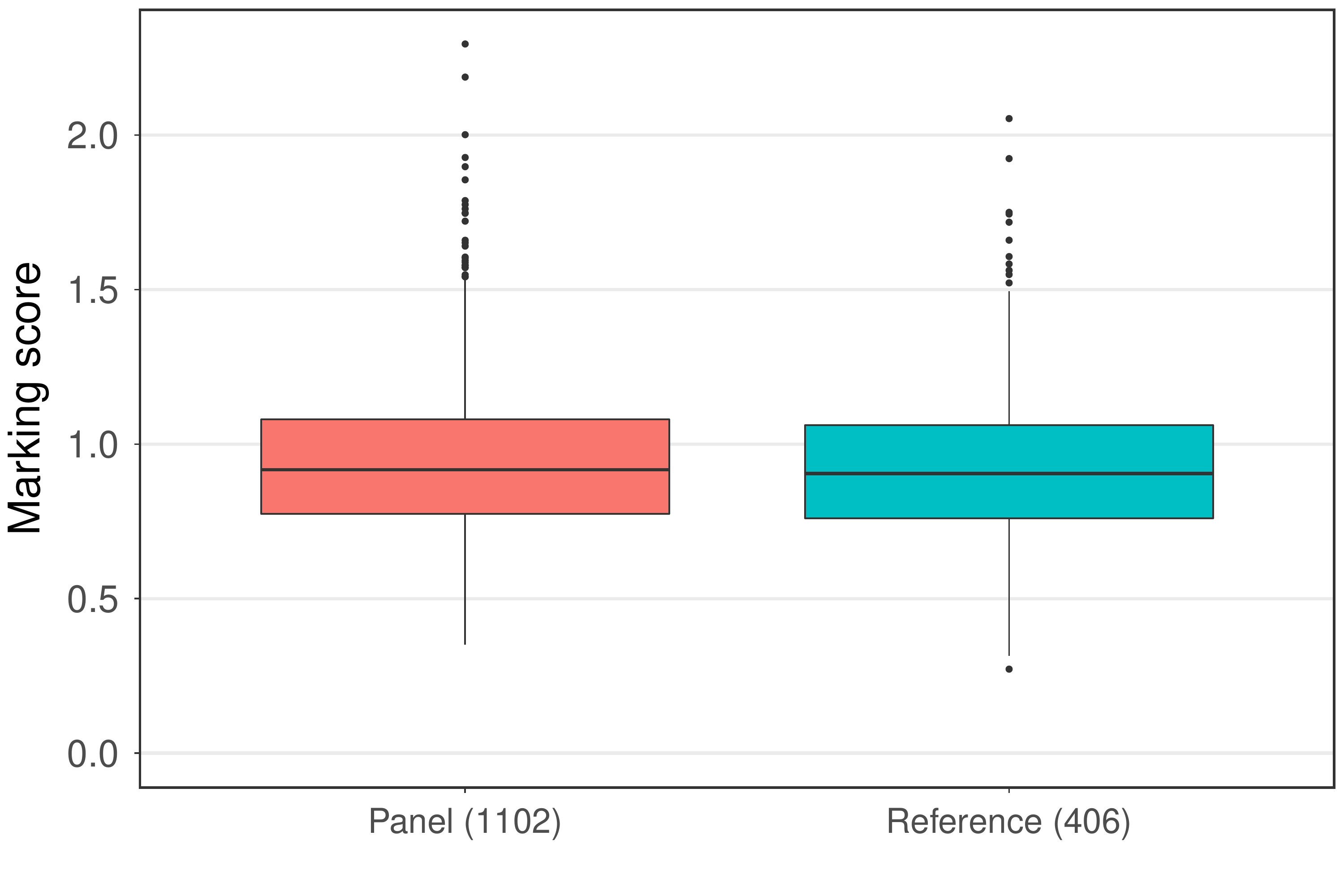}
	\caption{Distribution of marking scores for Artistic Gymnastics execution panel and reference judges.}
	\label{fig:pr}
\end{figure}

At each competition, execution judges are randomly selected from a set of accredited judges submitted by the national federations. In contrast, reference judges are hand-picked by the FIG,  and the additional power granted to them is based on the assumption that execution judges are sometimes incompetent or biased. To test this assumption, we compared the marking scores of the execution panel and reference judges. The results for artistic gymnastics are shown in Figure~\ref{fig:pr}\footnote{In Figure~\ref{fig:pr}, judges have at least one marking score per apparatus for which they evaluated gymnasts. A judge has two marking scores on a single apparatus when appearing on the regular execution panel and on the reference panel for different events.}. Although this is obvious by inspection, a two-sided Welch's $t$-test returned a $p$-value of 0.18 and we could not reject the null-hypothesis that both means are equal.

We ran similar tests for the other gymnastics disciplines, and in all instances reference judges are either statistically indistinguishable from the execution panel judges, or worse. Having additional judges selected by the FIG is an excellent idea because it increases the size of the panels, thus making them more robust. However, we strongly recommended that the FIG does not grant more power to reference judges. They are not better in aggregate, and the small size of the reference panels further increases the likelihood that the errors they make have greater consequences. The FIG Technical Coordinator has recently proposed the adoption of our recommendation.

\subsection{Gender discrepancies: women are more accurate judges than men}

\label{subsection:gender}

\begin{figure}[]
	\centering
	\includegraphics[width=\columnwidth]{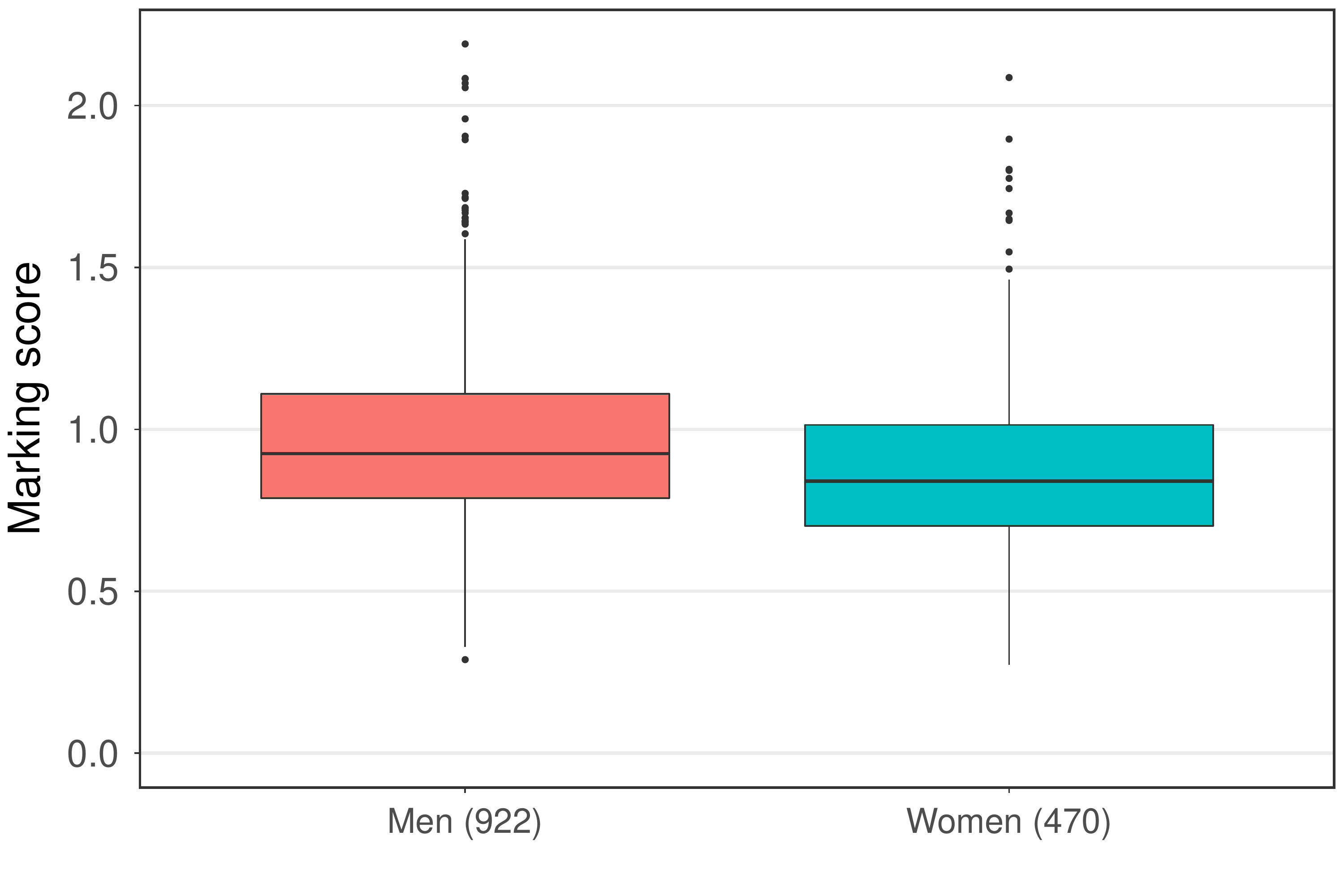} 
	\caption{Distribution of marking scores per gender in artistic gymnastics.}
	\label{fig:mwartistics}
\end{figure}

In artistic gymnastics, men apparatus are almost exclusively evaluated by men judges and women apparatus are almost exclusively evaluated by women judges. Figure~\ref{fig:boxplots:ag:all}, besides showing the differences between apparatus, also shows that the marking scores for women apparatus are lower than those of men apparatus. Figure~\ref{fig:mwartistics} formalizes this observation by directly comparing the marking scores of men and women judges in artistic gymnastics\footnote{In Figure~\ref{fig:mwartistics}, judges have one marking score per apparatus for which they evaluated gymnasts.}.

The average woman evaluation is $\approx 15\%$ better than the average man evaluation. More formally, we ran a one-sided Welch's $t$-test with the null-hypothesis that the mean of the marking scores of men is smaller than or equal to the mean marking score of women. We obtained a $p$-value of $10^{-15}$, leading to the rejection of the null-hypothesis.

A first hypothesis that can explain this difference is that in artistic gymnastics, men routines include ten elements, whereas women routines include eight elements. Furthermore, the formation and accreditation process is different for men and women judges. Men, who must judge six apparatus, receive less training than women, who must only judge four. Some men judges also have a (maybe unjustified) reputation of laissez-faire, which contrasts with the precision required from women judges. 

\begin{figure}[]
	\centering
	\includegraphics[width=\columnwidth]{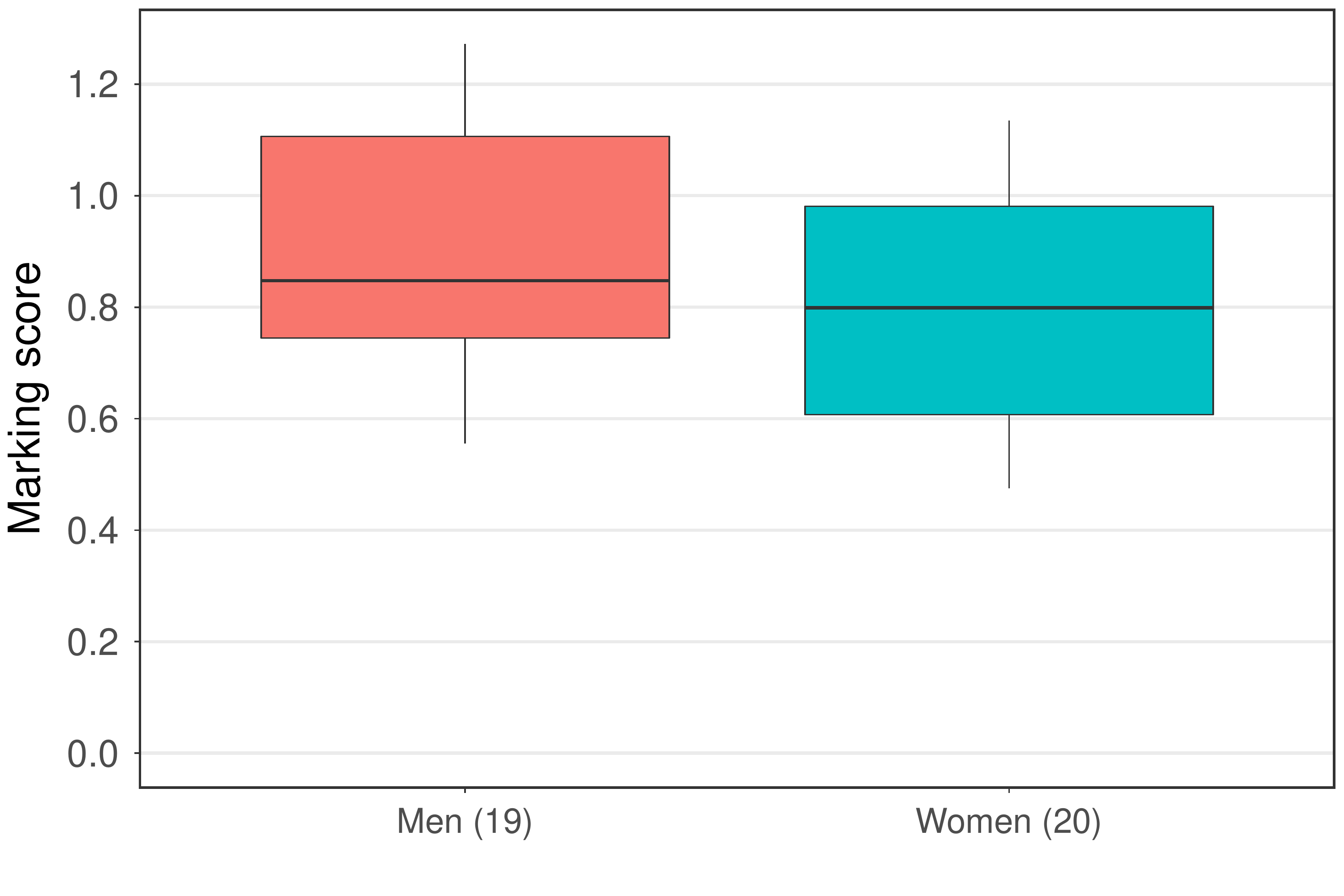} 
	\caption{Distribution of marking scores per gender in trampoline.}
	\label{fig:mwtrampoline}
\end{figure}

To obtain more insight, we compared women and men judges in trampoline, which has mixed judging panels as well as the same accreditation process and apparatus per gender. In other words,  men and women judges in trampoline receive the same training and execute the same judging tasks. The results are shown in Figure~\ref{fig:mwtrampoline}. The difference between gender observed in artistic gymnastics is less pronounced but remains in trampoline: women judge more accurately than men. We suspect that an important contributor of this judging gender discrepancy in gymnastics is the larger pool of women practicing the sport, which increases the likelihood of having more good women judges at the top of the pyramid since nearly all judges are former gymnasts from different levels. As an illustration, a 2007 Survey from USA Gymnastics reported four times more women gymnasts than men gymnasts in the USA \cite{USA2007survey}. A 2004 report from the ministère de la Jeunesse, des Sports et de la Vie Associative reported a similar ratio in France \cite{France2004survey}. Accurate information on participation per gender is difficult to come by, but fragmentary results indicate a similar participation gender imbalance in trampoline \cite{Silva:2017}.

On a different note, we did not observe any mixed-gender bias in trampoline, i.e., judges are not biased in favor of same-gender athletes. This in opposition to other sports such as handball where gender bias by referees led to transgressive behaviors \cite{Souchon:2004}.

In light of our gender analysis, we recommended that the FIG and its technical committees thoroughly review their processes to select, train and evaluate men judges in artistic gymnastics and trampoline. The marking score we developed provides valuable help for this task.

%% file: conclusion.tex
\section{Conclusions and limitations}
\label{sec:conclusion}

We put the evaluation of international gymnastics judges on a strong mathematical footing using robust yet simple tools. This has led to a better assessment of current judges, and will improve judging in the future. It is clear that there are significant differences between the best and the worst judges; this in itself is not surprising, but we can now quantify this much more precisely than in the past.

Our main contribution is a marking score that evaluates the accuracy of the marks given by judges. The marking score can be used across disciplines, apparatus and competitions. Its calculation is based on the intrinsic judging error variability estimated from prior data. Since athletes improve, and since Codes of Points are revised every four years, this intrinsic variability can and should be calibrated at the beginning of every Olympic cycle with data from the previous cycle. We calibrated our model using 2013--2016 data, and it should be recalibrated after the 2020 Tokyo Summer Olympics.

The FIG can use the marking score to assign the best judges to the most important competitions. The marking score is also the central piece of our outlier detection technique highlighting evaluations far above or below what is expected from each judge. The marking score and outlier detection work in tandem: the more accurate a judge is in the long-term, the harder it is for that judge to cheat without being caught due to a low outlier threshold detection. 

The FIG classifies international gymnastics judges in four categories: Category 1, 2, 3 and 4. Only judges with a Category~1 brevet can be assigned to major international competitions. The classification is based on theoretical and practical examinations, with increasingly stringent thresholds for the higher categories. As an example, in men's artistic gymnastics \cite{Artistic:2017MenRules} the theoretical examination for the execution component consists in the evaluation of 30 routines, 5 per apparatus. Our statistical engine is much more precise than the FIG examinations because it tracks judges longitudinally in real conditions over thousands of evaluations. Our dataset is dominated by Category 1 judges, and even at this highest level it shows significant differences among judges.

\subsection{Limitations of our approach: relative evaluations and control scores}
\label{subsection:discussion}

The first limitation of our approach is that judges are compared with each other and not based on their objective performance. An apparatus with only outstanding judges will trivially have half of them with a marking score below the median, and the same is true of an apparatus with only atrocious judges. From discussions with the FIG, no apparatus or discipline has the luxury of having only outstanding judges. We therefore proposed qualitative thresholds based on the fact that most judges are good, and a reward-based approach for the very best ones.

The second limitation of our approach is its dependence on accurate control scores.
Even though the Codes of Points are very objective in theory, in practice we must work with approximations of the true performance level, which remains unknown. This has implications for evaluating judges live during competitions and for training our model.

During competitions, quick feedback is necessary, and we approximate the control score with the median judging mark. Relying on the median for a single performance or a small event such as an Olympic final can be misleading. A high marking score for a specific performance is not necessarily an indicator of a judging error but can also mean that the judge is accurate but out of consensus with the other inaccurate judges. The FIG typically relies on observers like outside panels and superior juries to obtain quick feedback during competitions. We do not report detailed results here, but our analysis shows that like for reference judges, these observers are in the aggregate equal or worse than regular panel judges, and giving them additional power is dangerous.
Discrepancies between this outside panel and the regular panel should be viewed with circumspection. The best the FIG can do in this circumstance is to add these outside marks to the regular panel to increase its robustness until a more accurate control score is available post-competition.

The Technical Committee (TC) of each discipline calculates control scores post-competition using video reviews. Each TC uses a different number of members, ranging from two to seven, to evaluate each performance. Furthermore, each TC uses a different aggregation technique: sometimes members verbally agree on a score and other times they take the average. Even with video review, the FIG cannot guarantee the accuracy and unbiasedness of the TC members: some of them might be friends with judges they are evaluating and know what marks they initially gave. We therefore suggested clear guidelines for the calculation of the control scores post-competition to make them as robust as possible in the future. This is paramount to guarantee the accuracy of our approach on a per routine and per competition basis.

When we trained our model using 2013--2016 data, the FIG did not have control scores for every performance, and could not tell us under what conditions the available control scores had been derived. For this reason, we trained our model using the median of all the marks at our disposal. Considering the size of our dataset, this provides an excellent approximation of the intrinsic judging error variability.
However, like for live evaluations during competitions, retrospective judge evaluations during the 2013--2016 Olympic Cycle must be interpreted cautiously. While a longitudinal evaluation provides a very accurate view of the judges's performance, a bad evaluation for a specific event might indicate that the judge was accurate but out of consensus with other inaccurate judges. More precise control scores obtained by video review must once again be provided to settle the matter.

%% file: acknowledgements.tex
\section*{Acknowledgments}

This work is the result of fruitful interactions and discussions with the other project partners. We would like to thank Nicolas Buompane, Steve Butcher, Les Fairbrother, André Gueisbuhler, Sylvie Martinet and Rui Vinagre from the FIG, Benoit Cosandier, Jose Morato, Christophe Pittet, Pascal Rossier and Fabien Voumard from Longines, and Pascal Felber, Christopher Klahn, Rolf Klappert and Claudia Nash from the Université de Neuchâtel. This work was partly funded by Longines. A preliminary version of this work was presented at the 2017 MIT Sloan Sports Analytics Conference.